\documentclass{aa}
\usepackage{amssymb,amsmath,amsfonts}
\usepackage{graphicx}
\usepackage{color}
\usepackage{ulem}             
\usepackage{multicol}
\usepackage{caption}
\usepackage{pgf,tikz}
\usetikzlibrary{arrows}
\usetikzlibrary{trees}
\usetikzlibrary{snakes}
\usetikzlibrary{shapes}
\usetikzlibrary{matrix}
\usetikzlibrary{calc}
\usetikzlibrary{positioning}
\usetikzlibrary{chains}
\usetikzlibrary{shadows}

\tikzstyle{Point} = [fill, radius=0.08]
\tikzstyle{BigPoint} = [fill, radius=0.13]
\tikzstyle{Leaf} = [color = gray]
\tikzstyle{Line1} = [dashed]
\tikzstyle{Line2} = [dotted, ultra thick]\usepackage{colortbl}

\usepackage{xcolor}

\usepackage{amssymb}
\usepackage{latexsym}
\usepackage{pstricks}
\usepackage{pst-plot}
\usepackage{tabularx}
\usepackage{mathtools}
\usepackage{bm}
\usepackage{soul}
\usepackage{cancel}
\usepackage{hyperref}   

	\usetikzlibrary[patterns]

\newcommand\cZ{{\cal Z} }

\newcommand\default{{\textit{default}} }
\newcommand\highm{{\textit{high-mass}} }
\newcommand\final{{\textit{final}} }

\newcommand{\be}{\begin{equation}}
\newcommand{\ee}{\end{equation}}

\def\t{^\mathrm{T}}
\def\O{\mathcal{O}}

\usepackage[varg]{txfonts}
\usepackage{natbib}
\usepackage{amsmath}

 {\everymath{\displaystyle\everymath{}}\array}%
 {\endarray}

\begin{document}

\title{Unbiasing the density of TTV-characterised sub-Neptunes 
}
\subtitle{Update of the mass-radius relationship of 34 Kepler planets. }
\titlerunning{Update of the mass-radius relationship of 34 Kepler planets}

\author{
A. Leleu$^{1}$, J.-B. Delisle$^{1}$, S. Udry$^{1}$, R. Mardling$^2$, M. Turbet$^3$, J. A. Egger$^4$, Y. Alibert$^{4}$, G. Chatel$^5$, P. Eggenberger$^{1}$ and M. Stalport$^{1}$}
\authorrunning{A. Leleu et al}

\institute{
$^1$ Observatoire de Gen\`eve, Universit\'e de Gen\`eve, Chemin Pegasi, 51, 1290 Versoix, Switzerland.\\
$^2$ School of Physics and Astronomy, Monash University, Victoria 3800, Australia.\\
$^3$ Laboratoire de M\'et\'eorologie Dynamique/IPSL, CNRS, Sorbonne Universit\'e, \'Ecole Normale Sup\'erieure, PSL Research University, \'Ecole Polytechnique, 75005 Paris, France.\\
$^4$ Physikalisches Institut, Universit\"at Bern, Gesellschaftsstr.\ 6, 3012 Bern, Switzerland.\\
$^5$ Disaitek, www.disaitek.ai.\\
}
\abstract
{
Transit Timing Variations (TTVs) can provide useful information on compact multi-planetary systems observed by transits, by putting constraints on the masses and eccentricities of the observed planets. This is especially helpful when the host star is not bright enough for radial velocity follow-up. However, in the past decades, numerous works have shown that TTV-characterised planets tend to have a lower densities than RV-characterised planets. Re-analysing 34 Kepler planets in the super-Earth to sub-Neptunes range using the RIVERS approach, we show that at least part of these discrepancies was due to the way transit timings were extracted from the light curve, which had a tendency to under-estimate the TTV amplitudes. We recover robust mass estimates (i.e. low prior dependency) for 23 of the planets. We compare these planets the RV-characterised population. A large fraction of these previously had a surprisingly low density now occupy a place of the mass-radius diagram much closer to the bulk of the known planets, although a slight shift toward lower densities remains, which could indicate that the compact multi-planetary systems characterised by TTVs are indeed composed of planets which are different from the bulk of the RV-characterised population. These results are especially important for obtaining an unbiased view of the compact multi-planetary systems detected by Kepler, TESS, and the upcoming PLATO mission.
}


\maketitle

\section{Introduction}

The most high-yielding technique for detecting exoplanets is the transit method, which is based on the fact that when a planet passes in front of a star, the flux received from that star decreases. It has been, is, and will be applied by several space missions such as CoRoT, Kepler/K2, TESS, and the upcoming PLATO mission, to try and detect planets in large areas of the sky. When a single planet orbits a single star, its orbit is periodic, which implies that the transit happens at a fixed time interval. This constraint is used to detect planets when their individual transits are too faint with respect to the noise of the data: using algorithms such as Boxed Least Squares \citep[BLS,][]{Kovacs2002}, the data-reduction pipelines of the transit survey missions fold each light curve over a large number of different periods and look for transits in the folded data \citep[][]{Jenkins2010,Jenkins2016}. This folding of the light curve increases the number of observations per phase, and therefore also the signal-to-noise ratio (SNR) of the transit.

As soon as two or more planets orbit around the same star, their orbits cease to be strictly periodic. In some cases, the gravitational interaction of planets can generate relatively short-term transit timing variations (TTVs): transits no longer occur at a fixed period \citep{Dobrovolskis1996,Agol2005}. The amplitude, frequencies, and overall shape of these TTVs depend on the orbital parameters and masses of the planets involved \citep[see e.g.][]{Lithwick2012,NeVo2014,AgolDeck2016}. As the planet--planet interactions that generate the TTVs typically occur on timescales that are longer than the orbital periods, space missions with longer baselines such as Kepler and PLATO are more likely to observe such effects. Over the last decade, several efforts have been made to estimate the TTVs of the Kepler objects of interest (KOIs) \citep{Mazeh2013,RoTho2015,Holczer2016,Kane2019}. 

TTVs are a gold mine for our understanding of planetary systems: they can constrain the existence of non-transiting planets, adding missing pieces to the architecture of the systems \citep{Xie2014,Zhu2018}, and allowing for a better comparison with synthetic planetary-system population studies \citep[see e.g.][]{Mordasini2009, Alibert2013,Mordasini2018,Coleman2019,Emsenhuber2021}. TTVs can also be used to constrain the masses of the planets involved \citep[see e.g.][]{Nesvorny2013}, and therefore their density, which ultimately provides constraints on their internal structure, as is the case for the Trappist-1 system \citep{Grimm2018,Agol2020}. Detection of individual dynamically active systems also provides valuable constraints on planetary system formation theory, as the current orbital state of a system can display markers of its evolution \citep[see e.g.][]{BaMo2013,Delisle2017}. Orbital interactions also impact the possible rotation state of the planets \citep{DeCoLeRo2017}, and therefore their atmospheres \citep{Leconte2015,Auclair2017}.

However, if TTVs with amplitudes comparable to (or greater than) the duration of the transit occur on a timescale comparable to (or shorter than) the mission duration, there is no unique period that will successfully stack the transits of the planet. 
Fitting an orbit with a fixed orbital period to a planet with TTVs therefore results in estimating a shallower, longer transit, biasing the determined planet radii toward lower values \citep{Garcia2011}. 
{This erroneous transit shape is then commonly used for a first estimation of the planet transit timings \citep[e.g.][]{Rowe2014,Rowe2015,Holczer2016}. In \cite{Rowe2014,Rowe2015}, 
the transit shape is updated, correcting for TTVs in a second fit of the transit timings only if they found significant TTVs in the first estimation. In \cite{Holczer2016}, 
they recompute the transit shape while looking for TTVs only if the signal-to-noise ratio of individual transit is above 10, which is not the case for a large part of the sub-Neptune population detected by the Kepler mission (only 1 planet out of the 34 we re-analyse in this paper would satisfy this criterion).
%
For both methods, the lower the SNR of individual transits, the harder it is to fit the individual transits in the light curve, sometimes resulting in fitting background noise instead. All these effect combined can lead to incorrect timings estimations.
}
For the smallest planets, the TTVs signal can be completely missed, resulting in wrong estimation of planetary parameter, or even the mistaking of a planet for a false positive
\citep{RIVERS1,RIVERS2}. The correct planet parameters can be recovered using the photo-dynamical model of the light curve \cite{RaHo2010}, where the planet-planet interactions are modelled to account for TTVs \citep[e.g. Kepler-223, Kepler-444, Kepler-138][]{Mills16,Mills2017,Almenara2018}. However, for shallow transits and large TTVs, these photo-dynamical fits will struggle to converge if not initialised very close to the solution. In \cite{RIVERS1} we show how to tackle this problem using neural networks.

Numerous studies \citep[e.g.][]{WuLi2013,WeissMarcy2014,MillsMazeh2017,HaLi2017,Cubillos2017} have discussed the difference in density between the planets characterised by TTVs and radial velocities (RV). In particular, \cite{HaLi2017} re-analysed the TTVs of over 140 Kepler Objects of Interest (KOI)
and showed that the sub-population of planets whose masses were estimated by TTVs are less dense than the sub-population of planets for which the masses were estimated through RV.
This can be partially due to the bias inherent to each method: RVs tend to detect heavier planets and transits larger ones. In addition, for planets in the Earth to the mini-Neptune range, thus far TTVs tend to estimate masses further away from the host star, where the planets might be less subject to atmospheric escape. It is also possible that the observed populations are intrinsically different, as planets characterised by TTVs are embedded in compact multi-planetary systems that could undergo different formation and migration mechanisms, which is not necessarily the case for RV planets \citep[e.g. missing planets in compact systems][]{Delisle2018}. However, part of this discrepancy might be due to the difficulty in recovering the transit timings of small planets.

In this paper, we explore the result of applying the Recognition of Interval Variations in Exoplanet Recovery Surveys (RIVERS) method to 15 systems that were previously characterised using pre-extracted transit timings. The RIVERS method, described in detail in \cite{RIVERS1}, consists in applying a neural network to recover a proxy for the transit timings of each planet, then a photo-dynamical fit of the lightcurve. We first describe the method in section \ref{sec:method}. We then describe and discuss the newly obtained masses, radii and eccentricities in section \ref{sec:results}. A discussion and the conclusions can be found in section \ref{sec:conclusion}.

\section{Method}
\label{sec:method}

 \begin{figure}[!ht]
\begin{center}
\includegraphics[width=0.49\textwidth]{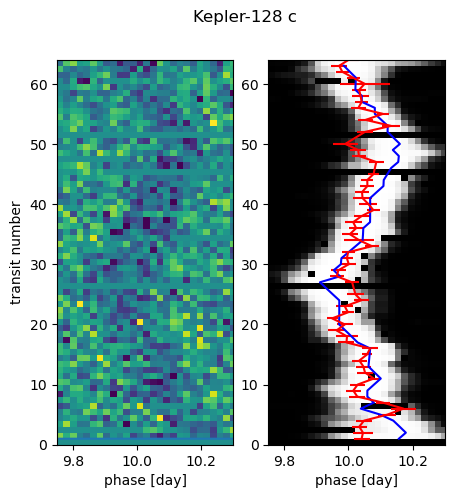}
\caption{\label{fig:Kepler128_RIVERS} Zoom on the track of Kepler-128 c in a river diagram with a folding period of 22.8030 days. The left panel shows the detrended data with a clipping at $3 \sigma$. The right panel shows the corresponding confidence matrix which is the output of the RIVERS.deep algorithm. Black indicates noise or missing data, white indicates the track of a planet. The track having the highest confidence is highlighted in blue. For comparison, the transit timings reported by \cite{Rowe2015} are shown in red. See Fig. \ref{fig:Kepler128_TTVs} for further comparison.   }
\end{center}
\end{figure}

\subsection{Selection of the targets}

In this paper we compare the output of a photodynamical fit of the lightcurve to the output of the fit of pre-extracted transit timings. To do so, we compare our results to those published in \cite{HaLi2017}. We focus on systems of (nearly-)resonant sub-Neptune objects exhibiting significant TTVs. The peak-to-peak amplitude of TTVs for each planet is estimated by taking the highest harmonics in the periodogram of the TTVs published by \cite{Rowe2015}.
 Out of the 55 systems studied by \cite{HaLi2017}, we select our systems as follows: systems composed only of planets with radius below $3.5 R_\oplus$ and with the sum of the peak-to-peak TTV amplitudes of all the planets above 40 minutes. We exclude systems of four or more (near-)resonant planets for simplicity. We also exclude Kepler-138 that was already studied by a photodynamical analysis \citep{Almenara2018}, as well as Kepler-29 that have been re-analysed by \cite{Vissapragada2020} with an additional transit from WIRC. We end up with 34 planets in 15 multi-planetary systems to analyse in this paper.

\subsection{Stellar parameters}
\label{sec:stars}

\renewcommand{\arraystretch}{1.5}

\begin{table*} 
\begin{small} 
\caption{Stellar parameters} 
\label{tab:stars} 
\centering 
\begin{tabular}{llllll} 
\hline 
Name & KIC & $T_{eff}$[K] & $M$[$M_\sun$] & $R$[$R_\sun$] & $\rho$[$\rho_\sun$]  \\ 
\hline\hline 
Kepler-23  & 11512246$^*$&$5828\pm {100}$&$1.078\pm {0.077}$&$1.548\pm {0.048}$&$0.291\pm {0.016}$ \\ 
Kepler-24  & 3231341&$6028\pm {116}$&$1.060\pm {0.069}$&$1.099\pm {0.039}$&$0.787\pm {0.098}$ \\ 
Kepler-26  & 9757613&$4124.0\pm {68.3}$&$0.593\pm {0.016}$&$0.595\pm {0.026}$&$2.75\pm {0.35}$ \\ 
Kepler-49  & 5364071&$4095.6\pm {74.7}$&$0.607\pm {0.014}$&$0.618\pm {0.020}$&$2.53\pm {0.21}$ \\ 
Kepler-28  & 6949607&$4499.3\pm {74.9}$&$0.684\pm {0.026}$&$0.664\pm {0.013}$&$2.32\pm {0.14}$ \\ 
Kepler-52  & 11754553&$4163.9\pm {62.2}$&$0.624\pm {0.017}$&$0.630\pm {0.025}$&$2.44\pm {0.30}$ \\ 
Kepler-54  & 7455287&$3853.9\pm {79.9}$&$0.518\pm {0.013}$&$0.522\pm {0.014}$&$3.61\pm {0.20}$ \\ 
Kepler-57  & 8564587&$5187.9\pm {90.8}$&$0.859\pm {0.049}$&$0.826\pm {0.019}$&$1.51\pm {0.15}$ \\ 
Kepler-58  & 4077526&$5747\pm {104}$&$0.972\pm {0.070}$&$0.982\pm {0.031}$&$1.01\pm {0.12}$ \\ 
Kepler-60  & 6768394&$6024\pm {113}$&$1.131\pm {0.082}$&$1.434\pm {0.039}$&$0.377\pm {0.045}$ \\ 
Kepler-85  & 8950568&$5505.2\pm {89.4}$&$0.928\pm {0.044}$&$0.875\pm {0.021}$&$1.38\pm {0.11}$ \\ 
Kepler-128  & 8077137$^*$&$6072.0\pm {75.0}$&$1.184\pm {0.074}$&$1.659\pm {0.038}$&$0.2591\pm {0.0048}$ \\ 
Kepler-176  & 8037145&$5139.2\pm {93.1}$&$0.847\pm {0.043}$&$0.800\pm {0.016}$&$1.64\pm {0.14}$ \\ 
Kepler-305  & 5219234&$5090\pm {101}$&$0.827\pm {0.046}$&$0.792\pm {0.025}$&$1.65\pm {0.17}$ \\ 
Kepler-345  & 9412760&$4722\pm {118}$&$0.786\pm {0.029}$&$0.819\pm {0.047}$&$1.39\pm {0.19}$ \\ 
\hline 
\end{tabular} 
\tablefoot{  
Stellar parameters from \cite{Berger2020}. $^*$ indicates that the parameters were updated, see sec. \ref{sec:stars}}
\end{small} 
\end{table*}

The stellar parameters reported in Table \ref{tab:stars} are taken from \cite{Berger2020} except for the two targets indicated by an asterisk. The stellar parameters of these two stars (KIC 11512246 and KIC 8077137) could indeed be refined thanks to the asteroseismic analysis of \cite{Huber2013}.




\subsection{Extraction of transit-timing proxy using RIVERS.deep}
\label{sec:RIVERSdeep}

The RIVERS.deep method, introduced in detail in \cite{RIVERS1}, is based on the recognition of the track of a planet in a river diagram \citep{Carter2012}. Figure \ref{fig:Kepler128_RIVERS} shows the example of Kepler-128 c, zoomed on the track of the planet. A river diagram displays the light curve in a 2-D matrix where each row shows one transit of the planet. The bottom row displays the first orbital period of the planet (22.8030 day in the case of Kepler-128 c), the second row displays the following orbital period, etc. The color code represents the normalised flux.

The RIVERS.deep algorithm takes this 2D array as input and produces two outputs: (1) {\bf A confidence matrix:} an array of the same size as the input containing for each pixel the confidence that this pixel belongs to a transit. This task is performed by the `semantic segmentation' (pixel-level vetting) subnetwork \citep{jegou2017one}. (2) {\bf A global prediction:} a value between 0 and 1 which quantifies the model confidence that the output of the semantic segmentation module is due to the presence of a planet. This task is performed by the classification subnetwork.

For this paper we already know the existence of the planets, so we only use the output of the semantic segmentation. The right panel of Fig. \ref{fig:Kepler128_RIVERS} shows a zoom on the track of the planet in the confidence matrix. The blue curve highlights the path of highest confidence. The red errorbars show the transit timings extracted by \cite{Rowe2015}. The TTVs associated with the red curve appears to be of lower amplitude than what we recover using RIVERS.deep. This is discussed further in sec. \ref{sec:amplitude}.

%
%
%
%

\subsection{Data pre-processing}
\label{sec:preprocess}
For each star, the raw PDCSAP flux is downloaded using the {\ttfamily lightkurve}\footnote{https://docs.lightkurve.org/} package. Long-cadence and short-cadence data is downloaded and pre-processed separately. The pre-processing is as follows: we start by checking for gaps longer than 2.5 h. Such gaps were commonly produced by the monthly data downlinks. After repointing the spacecraft, there was usually a photometric offset produced due to thermal changes in the telescope. We therefore removed all data points 6 hours prior, and 12 hours after such an interruption. We then created a copy of each lightcurve in which we removed in-transit data for all the known planets in the system {(a window of three time the transit duration centered on the transit timing predicted by the output of RIVERS.deep)}, and applied a B-spline detrending on the remaining data using {\ttfamily keplersplinev2}\footnote{https://github.com/avanderburg/keplersplinev2}. We used the `choosekeplersplinev2' function, forcing the timescale $\tau$ to remain between 3 times the longest transit duration of the planets in the system, and 1.5 days. The best $\tau$ (the one minimising the Bayesian Information Criterion) was then saved for use during the photo-dynamical fit (see section \ref{sec:photodyn}). We then checked the mean value of the detrended lightcurve within a sliding window of 5 hours of width. If this mean value departs from 1 by more than one time the standard deviation of the light curve for the long cadence data (resp. a third of the standard deviation for the short cadence data), it implies that the local behaviour of the lightcurve cannot be modelled by the B-splines. In this case, we flag all data point that are within the 5-hour window $\pm \tau$. 
We then use the raw data from which only the data near the downlinks were removed, and we additionally remove the data points we just flagged. This is the raw data that will be used in section \ref{sec:photodyn}.

\subsection{Photo-dynamical fit of the light curve}
\label{sec:photodyn}

The fits of the light curves were performed using a similar setup to the one presented in \cite{RIVERS1}. For each system, we use the adaptive MCMC sampler samsam\footnote{\url{https://gitlab.unige.ch/Jean-Baptiste.Delisle/samsam}} \citep[see][]{Delisle2018}, which learns the covariance of the target distribution from previous samples to improve the subsequent sampling efficiency. The transit timings of the planets were modelled using the {\ttfamily TTVfast} algorithm \citep{DeAgHo2014}. The approximate initial conditions for the orbital elements and masses of these planets were obtained by a preliminary fit of the transit timings to the timing proxy obtained from the application of RIVERS.deep (sec. \ref{sec:RIVERSdeep}, blue curve in Fig. \ref{fig:Kepler128_RIVERS}). We model the transit of each planet with the {\ttfamily batman} package \citep{batman}. For the long-cadence data, we use a supersampling parameter set to $29.42$ mins to account for the long exposure of the dataset. The effective temperature, ${\log g}$, and metallicity of the star (see sec. \ref{sec:stars}) were used to compute the quadratic limb-darkening coefficients $u_1$ and $u_2$ and their error bars were adapted to the Kepler spacecraft using {\ttfamily LDCU} \citep{Deline2022}. Based on the {\ttfamily limb-darkening} package \citep{EsJo2015}, {\ttfamily LDCU} uses two libraries of stellar atmosphere models ATLAS9 \citep{Kurucz1979} and PHOENIX \citep{Husser2013} to compute stellar intensity profiles for a given pass-band.

We work directly with the raw (non-normalized) fluxes obtained in sec. \ref{sec:preprocess}, and model them as
the product of the normalized transit model and a low-frequency component which
may account for both stellar variations and instrumental systematics.
This low-frequency component is modelled through a cubic B-spline
(see Appendix~\ref{sec:lmarg} for more details).
Our model for the raw flux $f$ is thus of the form
\begin{equation}
  f(\theta, \eta, t) = m(\theta, t) b(\eta, t),
\end{equation}
where $m$ is the normalized transit model with parameters $\theta$
and $b$ is the B-spline with parameters $\eta$.
We assume Gaussian white noise for the measurements error but
quadratically add a free jitter term $\sigma_\mathrm{jit.}$ to the measurements
error $\sigma_i$.
The full set of parameters of the model is $(\theta, \eta, \sigma_\mathrm{jit.})$
and the likelihood of the model is
\begin{align}
  \mathcal{L}(\theta, \eta, \sigma_\mathrm{jit.})
   & = p(y|\theta,\eta,\sigma_\mathrm{jit.})\nonumber                                         \\
   & = \frac{1}{\sqrt{|2\pi\Sigma|}} \exp\left(-\frac{1}{2} (y-f)\t \Sigma^{-1} (y-f) \right)
\end{align}
where $\Sigma = \mathrm{diag}(\sigma^2 + \sigma_\mathrm{jit.}^2)$
is the (diagonal) covariance matrix of the noise, and $y$ the observations.
We are mostly interested here in the transit related part of the model and
consider the B-spline parameters $\eta$ as nuisance parameters.
We thus analytically marginalize the likelihood over $\eta$ to obtain:
\begin{equation}
  \label{eq:lmarg}
  \mathcal{L}(\theta, \sigma_\mathrm{jit.})
  = p(y|\theta,\sigma_\mathrm{jit.})
  = \int p(y|\theta,\eta,\sigma_\mathrm{jit.}) p(\eta|\theta,\sigma_\mathrm{jit.}) \mathrm{d}\eta.
\end{equation}
The details of this procedure which allows for very efficient evaluations of the marginal likelihood
are explained in Appendix~\ref{sec:lmarg}.

\subsection{Masses, eccentricities and longitudes of periastron degeneracies}
\label{sec:ZZ2}

 \begin{figure}[!ht]
\begin{center}
\includegraphics[width=0.24\textwidth]{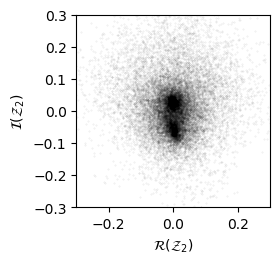}
\includegraphics[width=0.24\textwidth]{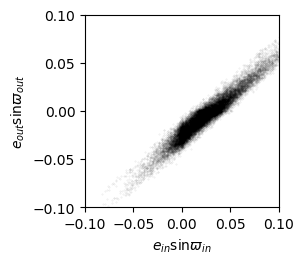}\\
\includegraphics[width=0.24\textwidth]{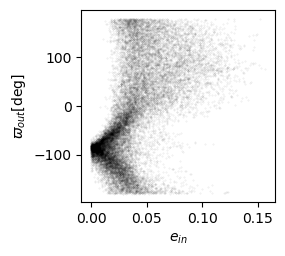}
\includegraphics[width=0.24\textwidth]{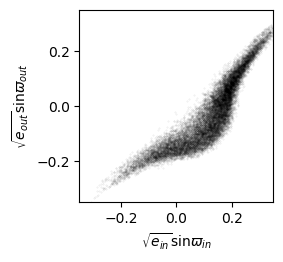}
\caption{\label{fig:core} Examples of correlation between parameters of the posterior of Kepler-345 (see section \ref{sec:results}). These correlation appears when $|\cZ_2|$ is poorly constrained. See Fig \ref{fig:post_ZZ2} for the full corner plots for each set of coordinates.}
\end{center}
\end{figure}

Taking $P_{in}$ (resp. $P_{out}$) to be the period of the inner (resp. outer) planet, these planets are close to two-planet mean motion resonance (MMR) when $P_{out}/P_{in} \sim (k+q)/k$, where $k$ and $q$ are integers, with $q$ the order of the resonance.
All the systems we study in this paper are close to first-order MMR (see section \ref{sec:eccres}), except Kepler-60 whose three planets are pairwise \textit{inside} first order MMRs and Kepler-26 which is near a second-order MMR ($q=2$). \cite{NeVo2016} describe the TTVs induced by first-order MMRs using an analytical model at first order in eccentricities, based on the 1-degree-of-freedom model proposed by \cite{HenLe1983}. The changes of variables from the 4-degree-of-freedom Hamiltonian of the two planets coplanar case to the 1-D model involves complex coordinates linked to the quantities:
\begin{equation}
\cZ =   e_{in} e^{i\varpi_{in}} + \frac{f_{31}}{ f_{27}} e_{out} e^{i\varpi_{out}}\, ,
\label{eq:Z} 
\end{equation}
and
\begin{equation}
\cZ_2 = e_{in} e^{i\varpi_{in}} - \frac{f_{27}}{ f_{31}} \sqrt{\frac{1}{\alpha}} \frac{m_{out}}{m_{in}}  e_{out} e^{i\varpi_{out}}\, . 
\label{eq:Z2}
\end{equation}
where $e_{in}$ is the eccentricity of the inner planet, $\varpi_{in}$ its longitude of periastron (resp. $e_{out}$ and $\varpi_{out}$ for the outer planet), and $f_{27}$ and $f_{31}$ are functions of the Laplace coefficients that depend on $k$ and $\alpha = a_{in}/a_{out}$, with $a_{in}$ (resp. $a_{out}$) the semi-major axis of the inner planet (resp. outer) and $m_{in}$ (resp. $m_{out}$) is the mass of the inner planet (resp. outer).
$\cZ$ evolves during the resonant motion and is linked to the TTV signal near \citep{Lithwick2012} or inside first-order MMRs \citep{NeVo2016}, while $\cZ_2$ has been shown to be a constant of motion at first order in eccentricities \citep{HenLe1983,NeVo2016}. 

\cite{HaLi2016} developed a model at second order in eccentricities, valid as long as the two planets are not too close to the resonance \citep[see ][Mardling 2022, in prep]{NeVo2016}. In this case, the TTV signal of near-resonant pair of planets can be split into three terms \citep{HaLi2016}: the first term, called `fundamental', which appears for pairs near first-order MMRs, is a sinusoidal signal whose frequency is associated with the super period:
\begin{equation}
P_{in,out} = ((k+1)/P_{out}-k/P_{in})^{-1}\, .
\end{equation}
The secondary term is a sinusoid with twice this frequency, and the third term is the high-frequency `chopping' signal. The chopping signal depends only on the masses, while the fundamental and secondary terms both depend on the masses, and $\cZ$. Observing only the fundamental term is hence not enough to disentangle $\cZ$ from the planetary masses: 
the chopping signal or the secondary term have to be observed as well. 

For planets closer to, or inside the MMR, recovering the masses requires observing either the secondary harmonics in the amplitude of libration of the resonant angle or the observation of a second, lower frequency signal \citep[][Mardling 2022, in prep]{NeVo2016}.

Good constraints on the masses hence generally translate to good constraints on the quantity $\cZ$, however the constant $\cZ_2$ needs to be determined to access the eccentricity and longitude of the periastron of the planets.
The full determination of the orbits hence requires recovering signals beyond the first order of eccentricities of the two-planet model\footnote{The absence of the dependence of the secondary term on $\cZ_2$ is an approximation, as shown by \cite{HaLi2016}. A well-constrained secondary term could hence help to constrain $\cZ_2$ as well.}.

As we will see in section \ref{sec:results}, in some cases the non-fundamental harmonics can be constrained well enough to estimate $\cZ_2$ as well. However, when $\cZ_2$ remains unconstrained, the posterior of the variables ($e_{in}$, $\varpi_{in}$ ,$e_{out}$, $\varpi_{out}$) is highly degenerate. 
Low-eccentricity solutions in the posterior distribution tend to correspond to small values of $|\cZ_2|$ and anti-aligned longitudes of periastra \citep[see Fig. 11 of][]{RIVERS1}, while higher-eccentricity solutions tend to correspond to larger values of $|\cZ_2|$ and to aligned longitudes of periastra.
Such a peculiar posterior shape is also found for resonant systems \citep[e.g.][]{Panichi2019,RIVERS1,RIVERS2}. In this case, a large value of $|\cZ_2|$ 
can lead to a precession of the longitude of the periastra of the planets, which can lead to a circulation of the resonant angles of the pair despite the fact that the orbit is formally resonant.

The dependence of the analytical TTV expression on $\cZ$ only, suggests that the degeneracies discussed above may be best explored via the coordinate $e_i \cos \varpi_i $ and $e_i \sin \varpi_i$ (which themselves are linear combinations of the real and imaginary parts of $\cZ$ and $\cZ_2$). Indeed, this choice produces elliptic-shaped correlations between these parameters, while other such as ($e_i$,$\varpi_i$) or ($\sqrt{e_i} \cos \varpi_i $,$\sqrt{e_i} \sin \varpi_i$) produce contorted correlations (see Fig. \ref{fig:core}) and can prevent our MCMC from properly exploring the parameter space.

\subsection{Priors}
\label{sec:priors}

We use wide, flat priors for the mean longitude, period, impact parameter, jitter, and ratio of the radius of the planet over the radius of the star $R_p / R_\star$. The stellar density and limb-darkening parameters have Gaussian priors.
 In order to test for the mass-eccentricity degeneracy inherent to pairs of planets near mean-motion resonances \citep{Boue2012,Lithwick2012}, \cite{HaLi2017} fitted the transits using different priors for masses and eccentricities. Their \default prior is log-uniform in planet masses
 and uniform in eccentricities. 
 Their \highm prior is uniform in planet masses 
 and log-uniform in eccentricities. 
 %
 We use the same choice of priors in order to be able to compare the posteriors of the photo-dynamical fit to their fit of pre-extracted transit timings. 
 In addition, we perform a third fit, using log-uniform mass prior and the \cite{Kipping2013} prior for the eccentricity: a $\beta$-distribution of parameters $\alpha=0.697$ and $\beta = 3.27$. The posterior associated with this set of prior is referred to as the \final posterior.


\section{Results}
\label{sec:results}

 The transit timings estimated for each planet, as well as 300 samples of the \final posterior for each system, can be found online.

\subsection{Pre-extracted transit timings or photo-dynamical analysis: effect on the mass-radius relationship of exoplanets}

\begin{figure}[!ht]
\begin{center}
\includegraphics[width=0.49\textwidth]{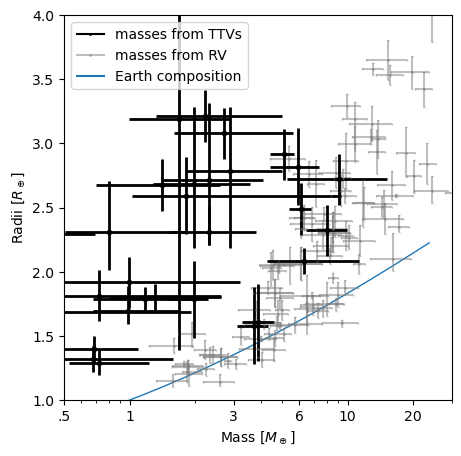}
\includegraphics[width=0.49\textwidth]{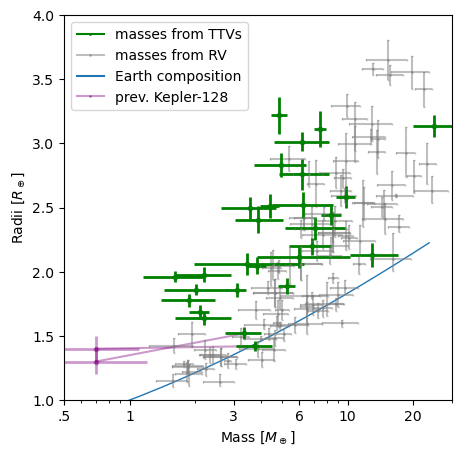}
\caption{\label{fig:MRdefault}  Posteriors of the mass-radius relation for a subset of the Kepler planetary systems studied in this paper. Black bars show the radii and masses resulting from the analysis of pre-extracted transit timings \citep{HaLi2017}. Green bars show the outcome of RIVERS (RIVERS.deep + the photo-dynamical fit). The priors on masses and eccentricities are set to the \default prior and identical in both studies. 
The blue line shows the earth's composition. The position of Kepler-128 from the top panel is also shown on the bottom panel in purple, with a purple line showing its shift in the mass-radius diagram.}
\end{center}
\end{figure}


In the introduction we explained that the use of pre-extracted transit timings \citep[e.g.][]{Rowe2014,Rowe2015,Holczer2016} might not be ideal when the SNR of individual transits (SNR$_i$, {that we define as the SNR reported by the Kepler team divided by the square root of the number of transits reported by the Kepler team}) is too low: adding an additional free parameter per transit might not fully recover the information from the light curve when individual transits are below the noise level. In this section   
%
%
we compare the posterior of the fit of pre-extracted transit timings to the posterior of a photo-dynamical fit. For this comparison, we use the \default set of priors for masses and eccentricities from \cite{HaLi2017}. We illustrate in Fig. \ref{fig:MRdefault} the difference in the posterior of the masses and radii of exoplanets depending on the use of pre-extract transit timings \citep[see][for the posterior of the masses and references therein for the radii]{HaLi2017} in black or photo-dynamical analysis (this paper) in green. The 5 most degenerate mass posteriors were removed from this plot to highlight the trend (see table \ref{tab:MR}). The bulk of the previous radius estimates had rather large uncertainties, making it hard to distinguish a clear trend in the shift of the radius of the planets. The masses however shift toward a larger value as a result of the photo-dynamical analysis. The photo-dynamical posterior of Kepler-128, which exhibits a shift from $\sim 0.7 M_\oplus$ for both planets to masses in the $3$ to $4$ $M_\oplus$ range, is highlighted in purple in Fig. \ref{fig:MRdefault}. Two causes were identified to explain such a difference: the pre-extracted transit timings yield lower-amplitude TTVs compared to those which are recovered with the photo-dynamical fit, and the photo-dynamical fit can constrain TTVs beyond the first harmonic, allowing the mass-eccentricity degeneracy to be broken \citep{Lithwick2012,HaLi2017}.

\subsubsection{Amplitude of recovered TTV signal}
\label{sec:amplitude}
 \begin{figure}[!ht]
\begin{center}
\includegraphics[width=0.49\textwidth]{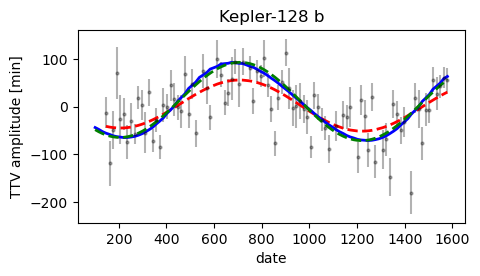}\\
\includegraphics[width=0.49\textwidth]{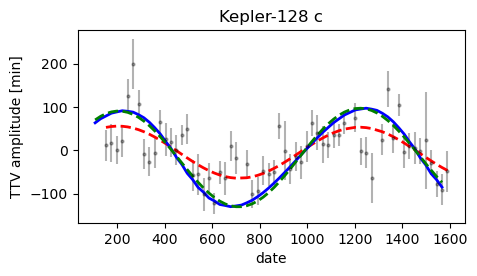}
\caption{\label{fig:Kepler128_TTVs} TTVs of the near-resonant pair Kepler-128b,c. Black bars show the pre-extracted transit timings from \cite{Rowe2015}, while blue bars show the best photodynamical fit. Dashed red and green curves show sinusoidal approximations for the 
pre-extracted transits and the photo-dynamical analysis respectively. }
\end{center}
\end{figure}

 \begin{figure}[!ht]
\begin{center}
\includegraphics[width=0.49\textwidth]{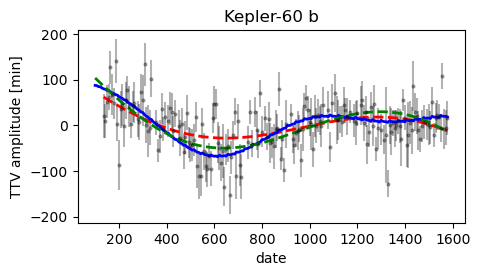}
\includegraphics[width=0.49\textwidth]{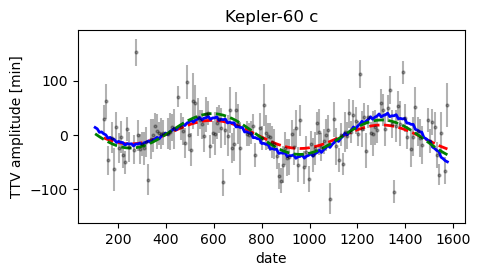}
\includegraphics[width=0.49\textwidth]{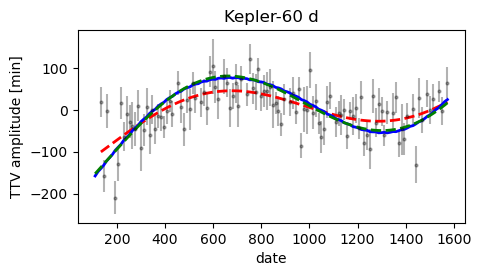}
\caption{\label{fig:Kepler60_TTVs} TTVs of the Laplace-resonant chain Kepler-60. See Fig \ref{fig:Kepler128_TTVs} for the description. }
\end{center}
\end{figure}

To illustrate the effect that pre-extracting the TTVs has on the recovered TTV amplitude, we begin with Kepler-128 (Fig. \ref{fig:Kepler128_TTVs}).
 The pre-extracted transit timings are shown in black with error bars \citep{Rowe2015}, while the curves going through the best-fit TTVs of the photo-dynamical model are shown in blue. Its main frequency is extracted. Then, a linear trend and a sinusoid of that frequency is fit to the pre-extracted transit timings in red (dashed) and to the TTVs of the best fit in green (dashed).
In this example, the sinusoidal approximations of the pre-extracted TTVs and of the photo-dynamical model show strong differences. The SNR$_i$ of both planets of Kepler-128 are rather small ($\sim 2.7$ and $3.1$) as a result, their pre-extracted transit timings appears to not be efficiently recovered. If we assume that the two-planet model is correct, it also appears that the errorbars of the pre-extracted transit timings are under-estimated. Another system, Kepler-60 is one of the rare system known in Laplace resonance and is hence of particular interest. With three known planets with SNR$_i$s in the range $1.68$ (Kepler-60 d) to $2.41$ (Kepler-60 c), we show the effect of the TTV pre-extraction in Fig. \ref{fig:Kepler60_TTVs}. {While the sinusoidal approximation may be less valid in this case, the first hundred days of Kepler-60d shows what might be the result of the initialisation for the search of TTVs: the individual transit timings are initialised on a fixed-period ephemeride, then allowed to vary. Some of the transit timings caught the real planet track (three points near $-150$ mins of TTV), while others (three points near $0$ mins of TTV) probably became trapped in a local minimum closer to $0$ mins TTVs, their initialisation. More generally, the pre-extracted TTVs of the three planets of Kepler-60, like the two planets of Kepler-128, shows numerous outliers, implying that the local search for each transit timings found noise structure that were preferred to the actual transit.}

\begin{figure}[!ht]
\begin{center}
\includegraphics[width=0.49\textwidth]{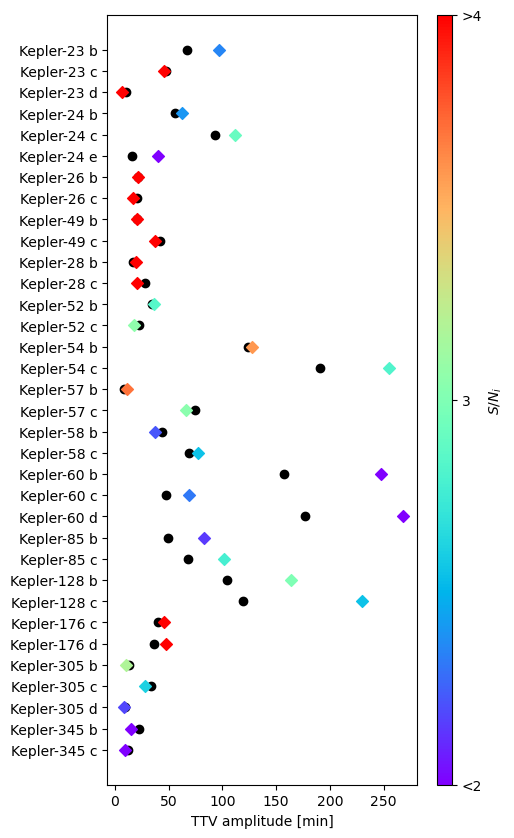}
\caption{\label{fig:TTV_amplitude} Peak-to-peak TTV amplitudes of the sinusoidal approximation of all of the planets that were analysed in this study. Black points are fits to the pre-extracted timings published by \cite{Rowe2015}, which for Kepler-128 and Kepler-60 correspond to the red-dashed curves in Figs.~\ref{fig:Kepler128_TTVs} and \ref{fig:Kepler60_TTVs}. The coloured diamonds show the peak-to-peak amplitude of the sinusoidal approximation of the best fit of the photo-dynamical model (the green dashed curves in Figs.~\ref{fig:Kepler128_TTVs} and \ref{fig:Kepler60_TTVs}). The colour indicates the SNR$_i$ of the planet. The agreement between pre-extracted timings and photo-dynamical fit is reduced for lower SNR$_i$. This is further highlighted in Fig. \ref{fig:TTV_amplitude_SNRi}. }
\end{center}
\end{figure}

\begin{figure}[!ht]
\begin{center}
\includegraphics[width=0.49\textwidth]{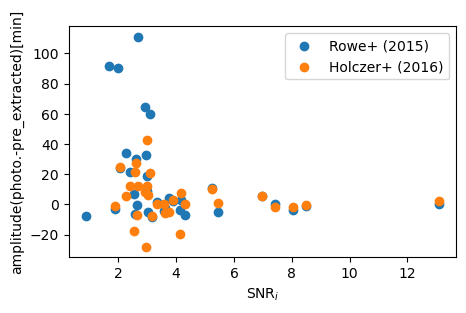}
\includegraphics[width=0.49\textwidth]{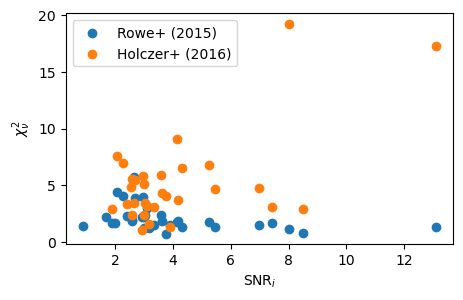}
\caption{\label{fig:TTV_amplitude_SNRi} Top: difference between the amplitude of the sinusoidal approximation of both the best solution of the photo-dynamical analysis and pre-extracted transit timings. The blue dots correspond to the timings from \cite{Rowe2015}, and in orange from \cite{Holczer2016}. Bottom: reduced chi-squared between the published transit timings and the best solution of the photo-dynamical fit. }
\end{center}
\end{figure}

The effect of the SNR$_i$ on the recovered TTV amplitude may be validated by an analysis of the whole dataset. Figure \ref{fig:TTV_amplitude} shows the difference between the peak-to-peak TTV amplitude for the sinusoid approximation and the pre-extracted transit timings (in black) and photo-dynamical analysis (coloured).  In all cases where SNR$_i \gtrsim 3.5$, the recovered amplitude of TTVs varies by a few minutes at most, while the amplitude can differ by several tens of minutes for lower SNR$_i$s. {This effect is further highlighted in Fig. \ref{fig:TTV_amplitude_SNRi}. In that figure, we show the pre-extracted TTVs from \cite{Rowe2015} that are used by \cite{HaLi2017}. We also show the timings of \cite{Holczer2016} that were extracted by a different method. The top panel shows the difference in amplitude of the sinusoidal approximation, while the bottom panel shows the reduced chi-squared ($\chi^2_\nu$) of the published transit timings and their error with respect to the timings of our best photodynamical fit (blue curves in Figs. \ref{fig:Kepler128_TTVs} and \ref{fig:Kepler60_TTVs}). The two databases display different behaviours: the timings from \cite{Rowe2015} appear to miss most of the large amplitude TTVs for low SNR$_i$, but have a relatively lower $\chi^2_\nu$, while the timings from \cite{Holczer2016} appear to have a relatively better estimation of the overall TTV amplitudes, but with a relatively larger fraction of outliers (larger $\chi^2_\nu$). Interestingly, even planets with relatively large SNR$_i$ (Kepler-176 c, SNR$_i=13.1$) have a large number of outliers. These differences can be explained by their different approaches: the method of \cite{Rowe2015} is `local': they first initialise the transit timings along a fixed-period ephemeride, then, if significant TTVs are observed, they update the transit shape and recompute the transit times. In some cases, they use two transit timing
measurements to linearly extrapolate an estimate of
the next transit time to initialize the fitter \citep{Rowe2014}. This result in successive updates of an initially flat TTV curve. \cite{Holczer2016} did a broader search: they systematically searched through a grid of timings around the expected transit
time. Each transit hence a better chance of recovering the correct timing regardless of the other transits, but a wider search also increase the risk of fitting background noise, which can explain the increased $\chi^2_\nu$. We note that three of the smallest SNR$_i$ (Kepler-345 b SNR$_i=0.89$, Kepler-60 b SNR$_i=1.98$ and Kepler-60 d SNR$_i=1.68$) are not in the \cite{Holczer2016} database.}

{ In order to recover robust planetary masses, both the main TTV amplitude and the lower amplitude, higher frequency signals are important: in the case of systems outside of MMR, the main TTV amplitude is directly linked to the estimated mass of the planets, while higher harmonics help to break the mass/eccentricity degeneracy, see sections \ref{sec:ZZ2} and \ref{sec:MEdeg}.}


\begin{table*} [ht!]
\begin{small} 
\caption{Final posterior value for 34 Kepler planets} 
\label{tab:MR} 
\centering 
\begin{tabular}{llllllll} 
\hline 
Planet &   P      & $R$          & $M$           & $\Delta_{M}$ &  $\rho$ & $e$ &  $\Delta_{e}$\\ 
       &  [$day$] &  [$R_\oplus$] & [$M_\oplus$]   &            &  [$\rho_\oplus$]  &  & \\ 
\hline\hline 
Kepler-23 b / KOI-168.03& $7.10$ & $1.638_{-0.047}^{+0.047}$ & $2.56_{-0.40}^{+0.43}$ & $1.16$&$0.578_{-0.079}^{+0.088}$&$0.017_{-0.014}^{+0.019}$ & $1.24$ \\ 
Kepler-23 c / KOI-168.01& $10.74$ & $3.005_{-0.074}^{+0.074}$ & $7.81_{-1.20}^{+1.32}$ & $1.18$&$0.286_{-0.037}^{+0.042}$&$0.021_{-0.014}^{+0.009}$ & $0.32$ \\ 
Kepler-23 d / KOI-168.02& $15.27$ & $2.206_{-0.057}^{+0.057}$ & $4.44_{-1.21}^{+1.30}$ & $0.54$&$0.41_{-0.11}^{+0.11}$&$0.010_{-0.008}^{+0.014}$ & $1.24$ \\ 
Kepler-24 b / KOI-1102.02& $8.14$ & $2.348_{-0.091}^{+0.092}$ & $11.14_{-1.93}^{+2.25}$ & $1.16$&$0.85_{-0.17}^{+0.20}$&$0.023_{-0.016}^{+0.014}$ & $0.44$ \\ 
Kepler-24 c / KOI-1102.01& $12.33$ & $2.514_{-0.098}^{+0.098}$ & $9.99_{-1.66}^{+2.01}$ & $1.13$&$0.62_{-0.13}^{+0.14}$&$0.014_{-0.011}^{+0.011}$ & $1.25$ \\ 
Kepler-24 e / KOI-1102.03& $18.99$ & $2.46_{-0.11}^{+0.11}$ & $[0.06,4.66]$ & $3.23$&$-$&$[0.01,0.07]$ & $1.40$ \\ 
Kepler-26 b / KOI-250.01& $12.28$ & $3.22_{-0.15}^{+0.15}$ & $4.85_{-0.42}^{+0.44}$ & $0.11$&$0.142_{-0.022}^{+0.023}$&$0.021_{-0.013}^{+0.021}$ & $0.81$ \\ 
Kepler-26 c / KOI-250.02& $17.25$ & $3.11_{-0.14}^{+0.14}$ & $7.48_{-0.48}^{+0.49}$ & $0.16$&$0.243_{-0.035}^{+0.036}$&$0.013_{-0.010}^{+0.013}$ & $1.24$ \\ 
Kepler-49 b / KOI-248.01& $7.20$ & $2.579_{-0.086}^{+0.087}$ & $9.77_{-0.95}^{+0.94}$ & $0.04$&$0.559_{-0.071}^{+0.071}$&$[0.00,0.06]$ & $1.30$ \\ 
Kepler-49 c / KOI-248.02& $10.91$ & $2.444_{-0.082}^{+0.083}$ & $8.38_{-0.89}^{+0.92}$ & $0.06$&$0.564_{-0.077}^{+0.079}$&$0.008_{-0.005}^{+0.023}$ & $0.93$ \\ 
Kepler-28 b / KOI-870.01& $5.91$ & $1.959_{-0.042}^{+0.043}$ & $1.63_{-0.40}^{+0.51}$ & $0.98$&$0.215_{-0.055}^{+0.068}$&$[0.00,0.08]$ & $1.71$ \\ 
Kepler-28 c / KOI-870.02& $8.99$ & $1.857_{-0.042}^{+0.042}$ & $2.06_{-0.52}^{+0.70}$ & $0.91$&$0.32_{-0.09}^{+0.11}$&$0.017_{-0.014}^{+0.023}$ & $0.62$ \\ 
Kepler-52 b / KOI-775.02& $7.88$ & $2.40_{-0.10}^{+0.10}$ & $[2.41,6.82]$ & $1.45$&$-$&$[0.08,0.20]$ & $1.45$ \\ 
Kepler-52 c / KOI-775.01& $16.38$ & $2.122_{-0.092}^{+0.093}$ & $[7.85,23.27]$ & $1.71$&$-$&$0.012_{-0.009}^{+0.018}$ & $1.23$ \\ 
Kepler-54 b / KOI-886.01& $8.01$ & $1.856_{-0.057}^{+0.057}$ & $3.09_{-0.31}^{+0.30}$ & $0.31$&$0.478_{-0.058}^{+0.057}$&$0.022_{-0.015}^{+0.020}$ & $0.50$ \\ 
Kepler-54 c / KOI-886.02& $12.07$ & $1.688_{-0.055}^{+0.054}$ & $2.10_{-0.21}^{+0.20}$ & $0.32$&$0.431_{-0.052}^{+0.053}$&$[0.00,0.05]$ & $1.26$ \\ 
Kepler-57 b / KOI-1270.01& $5.72$ & $3.135_{-0.088}^{+0.090}$ & $25.06_{-4.91}^{+5.16}$ & $0.49$&$0.81_{-0.17}^{+0.18}$&$0.0162_{-0.0021}^{+0.0024}$ & $0.37$ \\ 
Kepler-57 c / KOI-1270.02& $11.60$ & $2.196_{-0.065}^{+0.067}$ & $6.86_{-1.43}^{+1.52}$ & $0.48$&$0.64_{-0.15}^{+0.16}$&$0.0725_{-0.0062}^{+0.0070}$ & $0.43$ \\ 
Kepler-58 b / KOI-1336.01& $10.22$ & $2.113_{-0.082}^{+0.082}$ & $[3.10,32.41]$ & $4.64$&$-$&$0.040_{-0.031}^{+0.044}$ & $0.79$ \\ 
Kepler-58 c / KOI-1336.02& $15.57$ & $2.062_{-0.085}^{+0.086}$ & $[1.59,23.52]$ & $4.94$&$-$&$[0.00,0.10]$ & $1.26$ \\ 
Kepler-60 b / KOI-2086.01& $7.10$ & $1.889_{-0.061}^{+0.062}$ & $5.26_{-0.44}^{+0.45}$ & $0.08$&$0.77_{-0.10}^{+0.11}$&$[0.00,0.04]$ & $1.35$ \\ 
Kepler-60 c / KOI-2086.02& $8.90$ & $2.049_{-0.066}^{+0.066}$ & $3.84_{-0.40}^{+0.39}$ & $0.14$&$0.438_{-0.065}^{+0.066}$&$0.0390_{-0.0027}^{+0.0020}$ & $0.17$ \\ 
Kepler-60 d / KOI-2086.03& $11.90$ & $2.511_{-0.092}^{+0.093}$ & $4.40_{-0.44}^{+0.44}$ & $0.03$&$0.273_{-0.042}^{+0.044}$&$0.0020_{-0.0015}^{+0.0051}$ & $1.06$ \\ 
Kepler-85 b / KOI-2038.01& $8.30$ & $1.778_{-0.050}^{+0.050}$ & $1.84_{-0.47}^{+0.60}$ & $1.17$&$0.33_{-0.09}^{+0.11}$&$0.020_{-0.016}^{+0.028}$ & $1.24$ \\ 
Kepler-85 c / KOI-2038.02& $12.51$ & $1.978_{-0.057}^{+0.056}$ & $2.15_{-0.57}^{+0.73}$ & $1.16$&$0.277_{-0.077}^{+0.098}$&$0.027_{-0.014}^{+0.028}$ & $0.78$ \\ 
Kepler-128 b / KOI-274.01& $15.00$ & $1.421_{-0.040}^{+0.040}$ & $3.79_{-0.66}^{+0.76}$ & $0.74$&$1.31_{-0.23}^{+0.27}$&$[0.00,0.12]$ & $1.29$ \\ 
Kepler-128 c / KOI-274.02& $22.80$ & $1.521_{-0.047}^{+0.047}$ & $3.38_{-0.59}^{+0.67}$ & $0.74$&$0.95_{-0.17}^{+0.19}$&$0.037_{-0.030}^{+0.026}$ & $0.46$ \\ 
Kepler-176 c / KOI-520.01& $12.76$ & $2.281_{-0.051}^{+0.052}$ & $[0.56,8.25]$ & $6.99$&$-$&$[0.01,0.10]$ & $1.85$ \\ 
Kepler-176 d / KOI-520.03& $25.75$ & $2.354_{-0.061}^{+0.062}$ & $[0.80,8.34]$ & $5.92$&$-$&$[0.00,0.08]$ & $2.09$ \\ 
Kepler-305 b / KOI-1563.01& $5.49$ & $2.829_{-0.094}^{+0.094}$ & $[3.06,8.35]$ & $1.67$&$-$&$0.0050_{-0.0034}^{+0.0035}$ & $1.09$ \\ 
Kepler-305 c / KOI-1563.02& $8.29$ & $2.495_{-0.087}^{+0.088}$ & $[2.13,6.35]$ & $1.78$&$-$&$[0.00,0.01]$ & $1.34$ \\ 
Kepler-305 d / KOI-1563.04& $16.74$ & $2.76_{-0.12}^{+0.12}$ & $6.20_{-1.34}^{+1.76}$ & $1.20$&$0.296_{-0.072}^{+0.089}$&$[0.00,0.05]$ & $1.27$ \\ 
Kepler-345 b / KOI-1977.02& $7.42$ & $1.080_{-0.076}^{+0.076}$ & $[0.65,3.72]$ & $3.09$&$-$&$[0.00,0.06]$ & $1.43$ \\ 
Kepler-345 c / KOI-1977.01& $9.39$ & $2.03_{-0.13}^{+0.13}$ & $[1.32,8.52]$ & $2.27$&$-$&$[0.00,0.05]$ & $1.30$ \\ 
\hline 
\end{tabular} 
\tablefoot{  
Values with errorbars come from the \final posteriors and displays median values and the 0.16-0.84 quantile uncertainties. $\Delta_M$ and $\Delta_e$ are the degeneracy metrics defined by eq. (\ref{eq:deltaM}) for $\Delta_M$ and equivalently for $\Delta_e$. Whenever $\Delta_M$ (resp. $\Delta_e$) is above 1.3, the reported value for $M$ (resp. $e$) is replaced by the interval between the lowest 0.16 quantile across the three posteriors and the highest 0.84 quantile across the three posteriors to highlight the degeneracy.}
\end{small} 
\end{table*}

\subsubsection{Mitigation of the mass-eccentricity degeneracy}
\label{sec:MEdeg}

 \begin{figure*}[!ht]
\begin{center}
\includegraphics[width=0.60\textwidth]{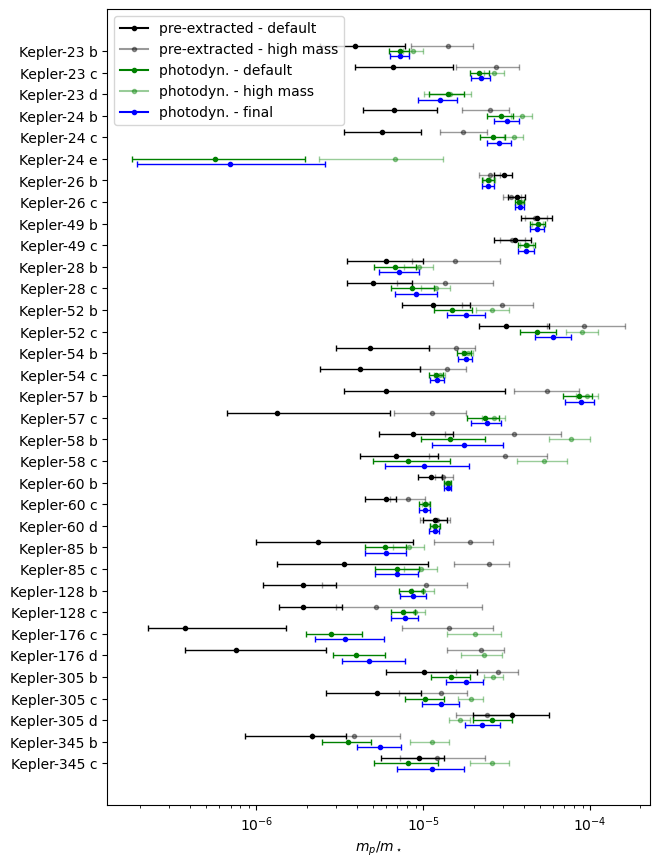}
\caption{\label{fig:mass_comp} In black and grey,  the \default and \highm posteriors of the fit of pre-extracted TTVs from \cite{Rowe2015} by \cite{HaLi2017}. In green and blue, are the posteriors from the RIVERS analysis. The dark green shows the \default posterior, the light green the \highm posterior and in blue the \final posterior. The full photo-dynamical analysis reduces the prior dependency, thereby increasing mass-estimate robustness, for most planets of the sample.
}
\end{center}
\end{figure*}

\begin{figure}[!ht]
\begin{center}
\includegraphics[width=0.49\textwidth]{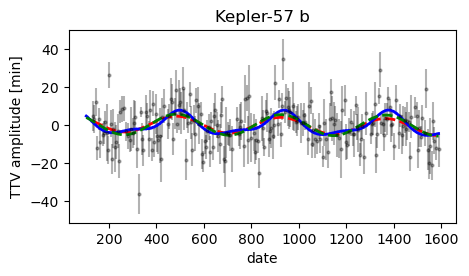}
\includegraphics[width=0.49\textwidth]{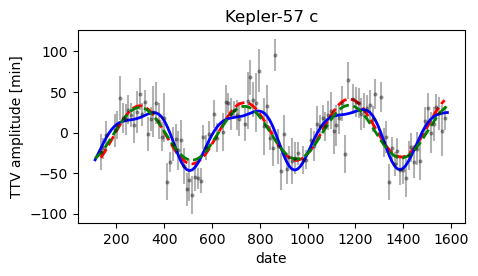}
\caption{\label{fig:Kepler57_TTVs} TTVs of the near-resonant system Kepler-57. See Fig \ref{fig:Kepler128_TTVs} for the description. }
\end{center}
\end{figure}


Following \cite{HaLi2017}, we explored the mass-eccentricity degeneracy intrinsic to systems close to but outside MMRs \citep{Lithwick2012}. The choice of priors is detailed in sec. \ref{sec:priors}. Figure \ref{fig:mass_comp} reports the mass posteriors of each planet resulting from the RIVERS analysis (RIVERS.deep + photo-dynamical analysis) in green and blue (this study), and the posteriors obtained by adjusting pre-extracted transit timings from \cite{Rowe2015} by \cite{HaLi2016} in black and grey. The \default prior tends to draw the posterior towards lower masses and larger eccentricities, while the \highm prior draws the posterior toward larger masses. The closer these two posteriors are, the less the posterior depends on the prior, and the more robust is the mass estimate. Figure \ref{fig:mass_comp} shows that the photo-dynamic analysis tends to significantly reduce the discrepancies between the posteriors, providing a more robust mass estimation. This is explained by the ability of the photo-dynamical analysis to better constrain the higher harmonics of the TTV signals; see section \ref{sec:MEdeg}. 
In addition to better constraining the amplitude of the dominant harmonic of the TTVs as shown in Fig. \ref{fig:mass_comp}, the photo-dynamical fit also better constrains the chopping or secondary term in several systems, reducing the mass-eccentricity degeneracy. Fig. \ref{fig:Kepler57_TTVs} shows the TTVs of Kepler-57, highlighting the recovery of a strong non-fundamental harmonic (see the blue curve). {The relative size of the TTV harmonics, computed from the analytical model of \cite{HaLi2016}, are shown and discussed in appendix \ref{ap:anaTTVs}.}

To provide our masses estimates, we chose to run a fit with the third set of priors, labelled as \final and shown in blue in Fig. \ref{fig:mass_comp},  which has the same log-uniform mass prior as the \cite{HaLi2017} \default prior. However, the \default prior is uniform in eccentricity. We chose the \final prior to being somewhat skewed toward lower eccentricities, as we deem it to be more realistic (see sec. \ref{sec:priors}). We then quantified the degeneracy impacting each mass estimate using the quantity 
\begin{equation}
\Delta_{M}=\max(\Delta_{M+},\Delta_{M-}).
\label{eq:deltaM}
\end{equation}
Here
\begin{equation}
\Delta_{M+} = \frac{ m_{high,0}-m_{final,0}}{m_{final,+\sigma}-m_{final,0}},
\label{eq:deltaMp}
\end{equation}
where $m_{high,0}$ is the maximum of the \default and \highm posterior medians, $m_{final,0}$ is the median of the \final posterior and $m_{final,+\sigma}$ is the $0.84$ quantile of the \final posterior, while
\begin{equation}
\Delta_{M-} = \frac{ m_{final,0}-m_{low,0}}{m_{final,0}-m_{final,-\sigma}},
\end{equation}
where $m_{low,0}$ is the minimum of the \default and \highm posterior medians, and $m_{final,-\sigma}$ is the $0.16$ quantile of the \final posterior. 
We then set an arbitrary threshold of $\Delta_{degen}=1.3$, below which we consider the masses to be constrained enough to be of use to the community. The masses and radii for all planets are reported in Table \ref{tab:MR}. For the planets which satisfy $\Delta_{degen} \leq 1.3$, the \final mass and density posterior is given. For planets where $\Delta_{degen} > 1.3$, an interval is given for the mass, with bounds corresponding to the lowest $0.16$ and highest $0.84$ quantiles of the three posteriors.
The eccentricity is treated in the same way, with the definition of $\Delta e$ reported in table \ref{tab:MR}. The reported radius comes from the \final posterior, although its value is always consistent across all three posteriors.

 \begin{figure}[!ht]
\begin{center}
\includegraphics[width=0.49\textwidth]{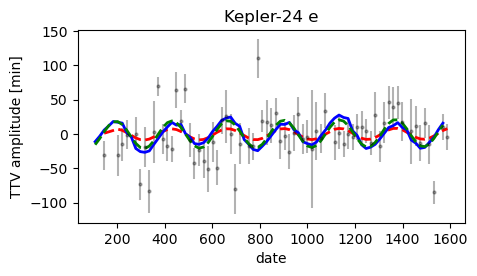}\\
\includegraphics[width=0.49\textwidth]{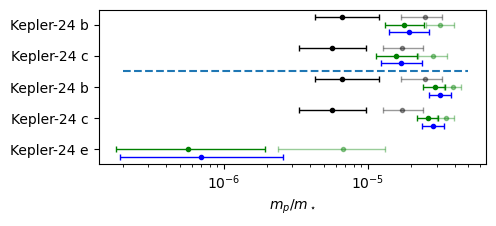}\\
\caption{\label{fig:Kepler-24} Top: TTVs of Kepler-24 e; see Fig \ref{fig:Kepler128_TTVs} for a description. The bottom panel shows the mass posteriors of the planets of Kepler-24, depending on the addition of Kepler-24 e to the model (see Fig. \ref{fig:mass_comp} for a description). Adding Kepler-24 e helps to better constrain the masses of  Kepler-24 b and  Kepler-24 c, despite having a highly degenerate mass itself.}
\end{center}
\end{figure}

\begin{figure*}[!ht]
\begin{center}
\includegraphics[width=0.6\textwidth]{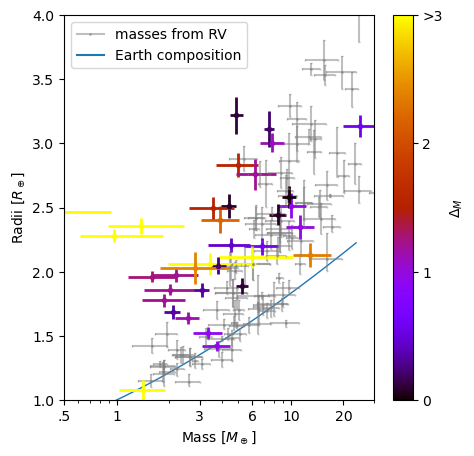}\\
\caption{\label{fig:MR_RV} Mass-radius relationship of the re-analysed set of Kepler planets with the \final posterior. The colour shows the degeneracy metric defined above eq. (\ref{eq:deltaM}): black-purple corresponds to well-constrained masses (low prior dependency) while red-yellow shows poorly constrained masses (high prior dependency). The grey background is the mass-radius relationship from RV-estimated masses. }
\end{center}
\end{figure*}

\cite{HaLi2017} considered a mass estimation robust if the median of the \highm posterior lay within the 0.16-0.84 quantile interval of the \default posterior. Out of the re-analysed systems, only Kepler-26 c, Kepler-49 b and c, and  Kepler-60 d were robust in their analysis (see the overlap of the black and grey errorbars in Fig. \ref{fig:mass_comp}). For these planets, we find very similar results across all our posteriors, implying that the two-priors test proposed by \cite{HaLi2017} is able to properly identify which masses are robust and which are not, regardless of the quality of the pre-extracted transit timings. In other words, poorly-determined transit timings did not `mimic' strongly constrained planetary masses in their analysis.  

The systems we analysed also partly overlap with those studied by \cite{Jontof2016}. To estimate the robustness of their solution, they also performed several tests including different eccentricity priors. Amongst the measurements labelled as `precise' are  Kepler-26, with $m_b=5.12^{+0.65}_{-0.61}$   and $m_c=6.20^{+0.65}_{-0.65} M_\oplus$, and Kepler-60: $m_b=4.19^{+0.56}_{-0.52}$,  
 $m_c=3.85^{+0.81}_{-0.81}$ and 
 $m_d=4.16^{+0.84}_{-0.75} M_\oplus$. These estimations are mostly $1-\sigma$ compatible with our results presented in Table \ref{tab:MR}, except for  Kepler-60b where the difference is of $\sim 2 \sigma$. Amongst their `less secure' solutions are Kepler 57 with $m_b=23.13^{+9.76}_{-7.64}$  and $m_c=5.68^{+2.55}_{-1.96} $ and Kepler-49 with $m_b=5.09^{+2.11}_{-1.9}$ and $m_c =3.28^{+1.45}_{-1.32} M_\oplus$. Although our result somewhat agrees for Kepler-57, we obtain a robust mass estimation that strongly disagrees with their solution for Kepler-49. Our results hence also agree with the robustness tests of \cite{Jontof2016}.
 
 From this, we draw two conclusions.
 \begin{itemize}
     \item Tests such as those presented in \cite{HaLi2017}, \cite{Jontof2016} or this paper are necessary to ensure the robustness of TTV-characterised masses.\\
     \item The recovery of the TTV signal (here using RIVERS.deep) and the photodynamical fit of the lightcurve can significantly increase the robustness of mass estimations.  
 \end{itemize}

 \subsubsection{Addition of a third planet}

Here we discuss the example of Kepler-24 with 4 known planets. \cite{HaLi2017} only considered Kepler-24b and c at $8.14$ and $12.33$ day. The inner planet at $4.24$ day is too far from any significant MMR with the other planets to have significant TTVs. Kepler-24 e, however, with an orbital period of $19.00$ days, has a period ratio of $\sim 1.54$ with Kepler-24c. Due to its small SNR$_i$ ($\approx 1.8$), no clear TTV signal is recognisable in the pre-extracted transit timings (see the top panel of Fig. \ref{fig:Kepler-24}). However, using RIVERS we were able to recover TTVs of $\sim 50$ minutes of peak-to-peak amplitude. The bottom panel of Fig. \ref{fig:Kepler-24} show the difference in the mass determination of Kepler-24 between the two- and three-planet model. In the three-planet model, Kepler-24e is unconstrained, but its TTVs help to break the degeneracy on the inner part of the system. A similar test was performed for Kepler-23: the addition of Kepler-23d helped to better constrain the whole system, also shifting the masses of Kepler-23b and Kepler-23c toward larger values while remaining $1 -2\sigma$ consistent with the two-planets solution.

\subsection{Masses, radii and densities}

\begin{figure*}[!ht]
\begin{center}
\includegraphics[width=0.99\textwidth]{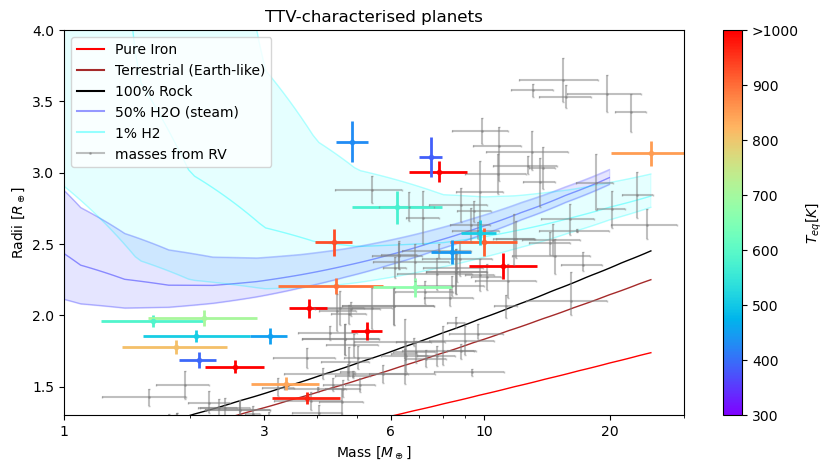}\\
\caption{\label{fig:MR_RV_Teq} Mass-radius relationship of the robust set of Kepler planets. The colour code shows the equilibrium temperature of the planet, taken from the exoplanet archive (computed assuming a bond albedo of 0.3). }
\end{center}
\end{figure*}

\begin{figure*}[!ht]
\begin{center}
\includegraphics[width=0.49\textwidth]{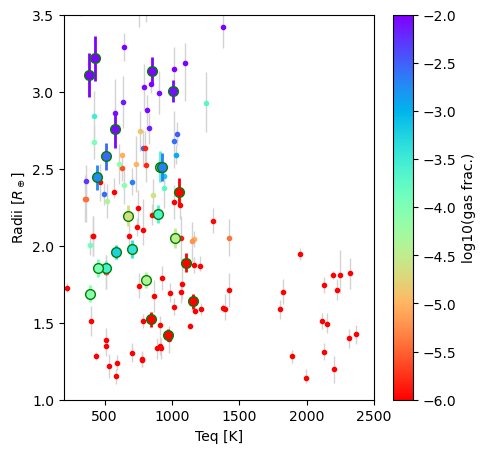}
\includegraphics[width=0.49\textwidth]{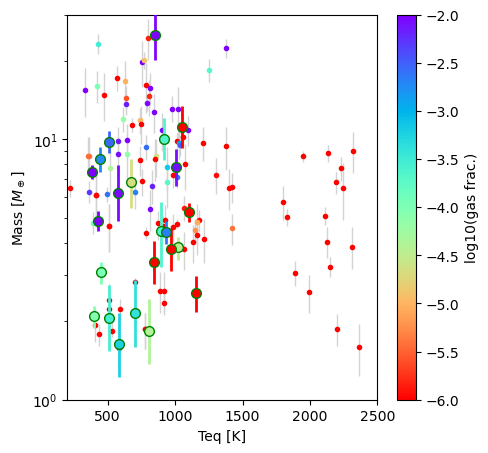}
\caption{\label{fig:R_Teq_gas} Radius (left) and mass (right) as function of equilibrium temperature of both the re-analysed set of Kepler planets (larger circles with green outline) and the RV-characterised planets. The colour code shows the modelled gas mass fraction of the planets in a logarithmic scale.}
\end{center}
\end{figure*}

Here we compare the mass-radius relationship of the sample of re-analysed Kepler planets to the samples of exoplanets for which the mass was estimated by the radial-velocity technique. We use the DACE database\footnote{https://dace.unige.ch/exoplanets/} \citep{Otegi2020} and select only the planets for which the mass uncertainty is below $50\%$ and the radius uncertainty below $30\%$. The mass-radius measurements of this population is shown in grey in Fig. \ref{fig:MR_RV}. The figure shows the posterior of the \final photo-dynamical fit, with the mass degeneracy indicator $\Delta_M$ colour-coded. The mass-radius measurements are also compared to various theoretical mass-radius relationships in Fig. \ref{fig:MR_RV_Teq}. This includes (1) pure solid interiors (pure iron, terrestrial Earth-like, pure MgSiO$_3$ rocky), taken from \citet{Zeng2016}; (2) terrestrial interiors with H$_2$/He envelopes, taken from \citet{Zeng19}; (3) terrestrial interiors with H$_2$O steam envelopes, taken from \citet{Aguichine2021}. The steam mass-radius relationships are the most appropriate here for water (with respect to liquid/icy interiors) given all planets of the sample are more irradiated than the runaway greenhouse limit \citep{Turbet2020}. Mass-radius relationships with H$_2$ and H$_2$O envelopes are arbitrarily ploted for various temperatures (500K, 700K, 1000K ; from the smallest to the largest radius) which are representative of the equilibrium temperatures of our sample of planets.

Firstly, we notice in Fig. \ref{fig:MR_RV} that the low-mass, large-radius population {amongst the re-analysed} planets (Kepler-24 e, Kepler-176 c and d) is in fact not robust. This is a reassuring result, as these planets should theoretically not be able to maintain a stable hydrogen envelope, given their low mass and the high irradiation they receive. These planets are indeed located in the part of the mass-radius diagram where the radius of a H$_2$-rich planet increases as the mass of the planet decreases (see Fig. \ref{fig:MR_RV_Teq}). This is symptomatic of the fact that the gravity of the planet is insufficient to guarantee the hydrostatic equilibrium of a H$_2$/He envelope.


Secondly, we notice that the robust TTV-characterized planets (this work) tend toward lower density than the RV-characterised planets, in particular the cluster of planets of masses between 2 and 3 $M_\oplus$ and radius below 2 $R_\oplus$. Most of the TTV-characterized planets lie above the pure rock (100$\%$ MgSiO$_3$, black curve in Fig. \ref{fig:MR_RV_Teq}), indicating that they must have a H$_2$-rich or a volatile (e.g., water) envelope \citep{Zeng19,Bean2021}. The mass-radius relationships alone do not lift the degeneracy between either of these two scenarios, at least for most planets of our sample. {It is still not clear whether the tendency toward lower density of TTV planets (compared to RV planets) is due to the sensibility of the RV method toward more massive planets (only plot RV planets for which the mass is known with better than 50\% are shown in Figs \ref{fig:MR_RV} and \ref{fig:MR_RV_Teq}, and low-mass planet have a tendency of having a lower precision on their mass}), or if it testifies distinct formation and evolution histories. We notice that the TTV planets have on average a lower equilibrium temperature than RV planets (Teq=$768 \pm 246$K for the TTVs samples and $1193 \pm 616$K for the RV sample), which could favor the stability of H$_2$ and/or H$_2$O envelopes, and thus increase the share of planets present in the large-radius peak of the so-called radius gap \citep{Fulton2017}. This is particularly relevant for the cluster of TTV planets near 2-3 $M_\oplus$ which have moderate equilibrium temperatures.


To quantitatively explore the differences in composition between the RV and TTV planet populations, we used a Bayesian inference method to characterise further the gas mass fraction of both populations. We hypothesize for this calculation that it is the gas envelope (H$_2$/He) that drives the observed variations in density between the planets. The full model is described in detail in \citep{Leleu2021} and based on \citep{Dorn2017}; see the appendix \ref{ap:internal_structure} for more details on the method. 

The results from our analysis are shown in Figure \ref{fig:R_Teq_gas}. It shows (Fig. \ref{fig:R_Teq_gas}, left panel) the gas mass fraction (more precisely the median of the posterior distribution of the gas mass fraction as derived from the Bayesian analysis) of the modelled planets from both samples in relation to their position in a radius and equilibrium temperature diagram. {In general, for both populations, the larger the radius, the larger the median of the gas mass fraction}. However, for equilibrium temperatures smaller than 1000\,K and radii between 1.5 and 2.5\,R$_\oplus$, the re-analysed set of Kepler planets shows on average {a much higher gas content. Indeed, the majority of RV-characterised planets have a gas mass fraction smaller than $10^{-6}$, whereas we see a significant fraction of Kepler planets with a gas mass fraction as large as $10^{-4}$}. The same can be observed when looking at the corresponding mass and equilibrium temperature diagram (Fig. \ref{fig:R_Teq_gas}, right panel), for masses below 3\,M$_\oplus$. This suggests that the equilibrium temperature difference between the RV and TTV planets cannot explain alone the differences in the gas content.

\subsection{Eccentricities and resonant states}
\label{sec:eccres}

\begin{table} 
\begin{small} 
\caption{Dynamical state of 19 Kepler pairs} 
\label{tab:GamZ} 
\centering 
\begin{tabular}{lllll} 
\hline 
Pair & $\frac{P_{out}}{P_{in}}$ &  $\Gamma '$ & $|\cZ|$ & $|\cZ_2|$ \\ 
\hline\hline 
Kepler-23 bc & $1.51$ & $-2.77_{-0.30}^{+0.45}$ & $0.0226_{-0.0027}^{+0.0032}$ & $0.065_{-0.032}^{+0.033}$ \\ 
Kepler-23 cd & $1.42$ & $-$ & $-$ & $-$ \\ 
Kepler-24 bc & $1.52$ & $-3.09_{-0.13}^{+0.21}$ & $0.0159_{-0.0017}^{+0.0026}$ & $0.026_{-0.012}^{+0.030}$ \\ 
Kepler-24 cd & $1.54$ & $-9.54_{-2.41}^{+2.55}$ & $0.036_{-0.016}^{+0.010}$ & $0.016_{-0.013}^{+0.012}$ \\ 
Kepler-26 bc & $1.41$ & $-$ & $-$ & $-$ \\ 
Kepler-49 bc & $1.52$ & $-2.90_{-0.18}^{+0.16}$ & $0.00433_{-9.1e-04}^{+8.9e-04}$ & $0.011_{-0.007}^{+0.049}$ \\ 
Kepler-28 bc & $1.52$ & $-5.19_{-1.86}^{+6.16}$ & $0.032_{-0.008}^{+0.013}$ & $0.050_{-0.019}^{+0.028}$ \\ 
Kepler-52 bc & $2.08$ & $-2.40_{-4.08}^{+9.40}$ & $0.071_{-0.013}^{+0.017}$ & $0.145_{-0.043}^{+0.072}$ \\ 
Kepler-54 bc & $1.51$ & $-1.69_{-0.08}^{+0.18}$ & $0.0142_{-0.0013}^{+0.0023}$ & $0.025_{-0.013}^{+0.033}$ \\ 
Kepler-57 bc & $2.03$ & $-5.73_{-0.37}^{+0.44}$ & $0.0515_{-0.0042}^{+0.0054}$ & $0.0322_{-0.0046}^{+0.0041}$ \\ 
Kepler-58 bc & $1.52$ & $-2.5_{-2.5}^{+13.8}$ & $0.043_{-0.017}^{+0.023}$ & $0.052_{-0.029}^{+0.057}$ \\ 
Kepler-60 bc & $1.25$ & $6.25_{-0.17}^{+0.17}$ & $0.02861_{-5.2e-04}^{+5.0e-04}$ & $0.0286_{-0.0052}^{+0.0066}$ \\ 
Kepler-60 cd & $1.34$ & $3.89_{-0.15}^{+0.14}$ & $0.02632_{-6.2e-04}^{+6.4e-04}$ & $0.0384_{-0.0046}^{+0.0041}$ \\ 
Kepler-85 bc & $1.51$ & $-2.68_{-0.35}^{+1.78}$ & $0.0179_{-0.0040}^{+0.0056}$ & $0.040_{-0.021}^{+0.070}$ \\ 
Kepler-128 bc & $1.51$ & $5.44_{-3.45}^{+6.47}$ & $0.0431_{-0.0063}^{+0.0091}$ & $0.060_{-0.025}^{+0.052}$ \\ 
Kepler-176 bc & $2.02$ & $-2.2_{-8.1}^{+26.9}$ & $0.041_{-0.017}^{+0.021}$ & $0.096_{-0.047}^{+0.059}$ \\ 
Kepler-305 bc & $1.51$ & $-4.23_{-0.53}^{+0.56}$ & $0.0042_{-0.0012}^{+0.0020}$ & $0.0057_{-0.0026}^{+0.0048}$ \\ 
Kepler-305 cd & $2.02$ & $-7.80_{-0.88}^{+0.66}$ & $0.0086_{-0.0059}^{+0.0090}$ & $0.018_{-0.014}^{+0.030}$ \\ 
Kepler-345 bc & $1.27$ & $-7.70_{-1.55}^{+1.27}$ & $0.0115_{-0.0046}^{+0.0049}$ & $0.024_{-0.014}^{+0.035}$ \\ 
\hline 
\end{tabular} 
\tablefoot{  
$\Gamma'$ is the resonant parameter discussed in sec. \ref{sec:eccres}. The complex quantities $\cZ$ and $\cZ_2$ are defined eq. (\ref{eq:Z}) and (\ref{eq:Z2}). The value reported are computed on the \final posterior ($\beta$-distribution as eccentricity prior), which somewhat under-estimate the uncertainties on $\cZ_2$ when it highly degenerates}
\end{small} 
\end{table}

The architecture of (nearly-)resonant systems contains information on their formation and evolution, such as migration in the proto-planetary disc \citep[e.g.][]{Nesvorny2021} and tidal evolution \citep[e.g.][]{Lee2013}. Table \ref{tab:GamZ} indicates the value of the parameter $\Gamma '$ for each pair of planets. $\Gamma '$ is the parameter of the one degree of freedom model of the first order MMRs, which describes the position of a pair of planets with respect to the closest first-order MMR\footnote{Numerous equivalent versions of the model exist in the literature, and we chose $\Gamma '$ from \citep{DePaHo2013} for consistency with \cite{RIVERS1,RIVERS2}.} \citep{HenLe1983,DePaHo2013}. The power of this 1-degree of freedom model is that it allows comparing pairs of planets of different orbital periods and masses with respect to any of the first order MMRs. The resonance formally appears for $\Gamma ' \geq 1.5$. Hence, if $\Gamma '<1.5$, the pair cannot be resonant (although its resonant angles can librate around a given value), and $\Gamma ' $ gives an estimate of the distance to the resonance.  For $\Gamma '>1.5$, an additional test is required to check if the system lies inside the resonance, which we will not describe here. For resonant systems, $\Gamma '$ describes how `deep' the pair is in the MMR. Breaking the degeneracy between the masses of the planets and $\cZ$ hence does not only provide good mass estimates for the planets involved, but can also provide valuable information about their resonant state, which can in turn be linked to their proto-planetary disc or the inner planet's internal structure.
Out of all the pairs studied, only Kepler-60 b and c and Kepler-60 c and d are inside the 2-body MMRs, forming a Laplace resonant chain \citep{Gozdziewski2016}. We obtained good constraints on  $\Gamma '$ for several of the pairs; the implications for the dissipative evolution of these systems will be the subject of a future study. Systems with large uncertainties on $\Gamma '$, such as Kepler-128, were checked to have all of their posterior outside of the resonance. 

We obtain robust eccentricity estimates for 19 planets. Often the errors remain quite large, with medians of a few percent and uncertainties of similar amplitude.  However some eccentricities are different from zero at more than $8\sigma$, such as Kepler-57 b and c and Kepler-60 c (see table \ref{tab:MR}). This implies that we were able to constrain the TTVs beyond the effects of first-order in eccentricities (see section \ref{sec:MEdeg}). This results in good constraints on $\cZ_2$ (see table \ref{tab:GamZ}). 

\section{Summary and Conclusion}
\label{sec:conclusion}


We re-analysed a sample of 34 Kepler planets in the super-Earth to mini-Neptune range in 15 multi-planetary systems. Most of these planets were known to have TTVs, with transit timings available in databases \cite{Rowe2015,Holczer2016}. These systems were previously characterised by fitting these pre-extracted transit timings \citep[e.g.][]{Jontof2016,HaLi2017}. Our analysis used the RIVERS method, which first estimates the transit timings of the planets using the RIVERS.deep algorithm (CNN-based image recognition, see section \ref{sec:RIVERSdeep}), then uses a photo-dynamical fit of the light curve (see section \ref{sec:photodyn}).

Firstly, we showed that the transit timings resulting from our analysis often differ by several tens of percent in amplitude from the published values, introducing a systematic bias in estimates of the associated planet masses. We have shown that this difference is strongly anti-correlated with the signal-to-noise ratio of individual transits of the planets, indicating that the classical approach of fitting the lightcurve, with individual transit timings as free parameters to recover the TTVs, gives poor results when the individual transit SNR is below $\sim 4$. 

Secondly, using the \default prior identical to the one used by \cite{HaLi2017}, 
we consistently recovered masses that were higher than those obtained by fitting pre-extracted transit timings. This difference, which can be more than $4\sigma$ as in the case of Kepler-128 \citep{HaLi2016}, is explained not only by the difference in TTV amplitude, but also by the capacity of the photo-dynamic analysis to recover additional harmonics (especially when short-cadence data is available), allowing one to break the mass-eccentricity degeneracy inherent to nearly-resonant pairs of planets. For the 3-planet systems Kepler-23 and Kepler 24, we also tried fitting only two planets; in both cases the 3-planet study yielded larger, better constrained masses for the two planets that were present in both analyses. {This highlights the model dependency inherent to the TTV method, and imply that non-transiting planets can also be responsible for an under-estimation of the masses of planets.}

Out of the 34 planets analysed, we robustly determine the mass of 23 planets (low prior dependency, see the robustness criterion sec. \ref{sec:MEdeg}), 13 of which have a precision on the mass better than 20\%.
Comparing the newly-characterised planets to the RV-characterised population from \citep{Otegi2020}, it appears that some of the TTV-characterised planets still have a lower density than their RV counterparts, which is in agreement with the robustly-characterised samples of \cite{HaLi2017}. In terms of internal structure, this lower density translates to a larger mass fraction of gas, as derived from internal structure modeling. 
Although the TTV-characterized planets we study here have on average a lower equilibrium temperature than RV-characterized planets (which would help stabilize a gas and/or volatile envelope), we notice that this alone cannot explain the observed differences in density. The difference could be due to the bias inherent to the method used to obtain the mass: higher masses make a planet easier to well characterise using RVs, while deeper transits allow for better transit timing estimates. Another explanation could be related to the fact that planetary systems characterised by TTVs are necessarily compact multi-planetary systems, hence dynamically `cold' (the different planets being by necessity almost coplanar). The difference in density could therefore be related to the formation and evolution history of the systems. 
More studies, comparing larger samples and correcting the effect of observational biases, are required in order to decipher the origin of the differences between the two populations. {We also leave to a future study the re-analysis of super-puff planets such as Kepler-79 \citep{Jontof-Hutter2014} \citep[found not robust by][]{HaLi2017}, and Kepler-51 \citep{Masuda2014,Libby-Roberts2020} \citep[found robust by][]{HaLi2017}.}


We constrained the eccentricities of most planets to a few percent at most, with errors of the same order.
Exceptions were Kepler-57 b and c and Kepler-60 c for which the analysis provided relatively precise estimates of the individual eccentricities, which in itself is of particular interest given that it requires adequate power in the non-dominant harmonics, and because it is especially difficult to measure such small eccentricities with radial velocities. Breaking the mass-eccentricity degeneracy also often allowed us to have a better view on the resonant state of the systems. For systems whose inner planet is far-enough from the star (typically $>10$days), this will allow us to constrain the local shape of the proto-planetary disc prior to its dispersal \citep[e.g.][]{Nesvorny2021}. For system whose inner planet is closer to the star, the observed resonant state can allow to constrain the tidal dissipation in the inner planets \citep[e.g.][]{Lee2013}.   

Finally, we want to stress that all of the robustly-characterised planets in this study occupy a `believable' position in the mass-radius diagram, while a single analysis of pre-extracted transit timings would often have resulted in a strong under-estimation of the planetary masses.
Since the quality of pre-extracted transit timings strongly correlate with the transits SNRs, hence the planetary radius, this bias mostly affects systems of small (typically sub-Neptune) planets, which can impede the characterisation of these systems in the Kepler, TESS and upcoming PLATO data.

Our results, combined with those of \cite{HaLi2017}, hence strongly advocate the use of several priors to explore the mass-eccentricity degeneracy, as well as the systematic use of photo-dynamical analysis to recover higher, smaller amplitudes harmonics of the TTVs signals. Since the long baseline of the Kepler data, near-polar observations of TESS, and PLATO allow for the detection of planets whose individual transits are below the noise level, recovering individual transits to initialise the photo-dynamical fit might be challenging. In this case, we have shown that the use of RIVERS.deep, based on the the recognition of the track of a planet in a river diagram using a neural network, could recover the transit timings of such planets. 


\begin{acknowledgements}
This work has been carried out within the framework of the National Centre of Competence in Research PlanetS supported by the Swiss National Science Foundation and benefited from the seed-funding program of the Technology Platform of PlanetS. The authors acknowledge the financial support of the SNSF.
\end{acknowledgements}

\bibliographystyle{aa}
\bibliography{biblio}

\begin{thebibliography}{86}
\expandafter\ifx\csname natexlab\endcsname\relax\def\natexlab#1{#1}\fi

\bibitem[{{Adibekyan} {et~al.}(2021){Adibekyan}, {Dorn}, {Sousa}, {Santos},
  {Bitsch}, {Israelian}, {Mordasini}, {Barros}, {Delgado Mena}, {Demangeon},
  {Faria}, {Figueira}, {Hakobyan}, {Oshagh}, {Soares}, {Kunitomo}, {Takeda},
  {Jofr{\'e}}, {Petrucci}, \& {Martioli}}]{Adibekyan2021}
{Adibekyan}, V., {Dorn}, C., {Sousa}, S.~G., {et~al.} 2021, Science, 374, 330

\bibitem[{{Agol} \& {Deck}(2016)}]{AgolDeck2016}
{Agol}, E. \& {Deck}, K. 2016, \apj, 818, 177

\bibitem[{{Agol} {et~al.}(2020){Agol}, {Dorn}, {Grimm}, {Turbet}, {Ducrot},
  {Delrez}, {Gillon}, {Demory}, {Burdanov}, {Barkaoui}, {Benkhaldoun},
  {Bolmont}, {Burgasser}, {Carey}, {de Wit}, {Fabrycky}, {Foreman-Mackey},
  {Haldemann}, {Hernandez}, {Ingalls}, {Jehin}, {Langford}, {Leconte},
  {Lederer}, {Luger}, {Malhotra}, {Meadows}, {Morris}, {Pozuelos}, {Queloz},
  {Raymond}, {Selsis}, {Sestovic}, {Triaud}, \& {Van Grootel}}]{Agol2020}
{Agol}, E., {Dorn}, C., {Grimm}, S.~L., {et~al.} 2020, arXiv e-prints,
  arXiv:2010.01074

\bibitem[{{Agol} {et~al.}(2005){Agol}, {Steffen}, {Sari}, \&
  {Clarkson}}]{Agol2005}
{Agol}, E., {Steffen}, J., {Sari}, R., \& {Clarkson}, W. 2005, \mnras, 359, 567

\bibitem[{{Aguichine} {et~al.}(2021){Aguichine}, {Mousis}, {Deleuil}, \&
  {Marcq}}]{Aguichine2021}
{Aguichine}, A., {Mousis}, O., {Deleuil}, M., \& {Marcq}, E. 2021, \apj, 914,
  84

\bibitem[{{Alibert} {et~al.}(2013){Alibert}, {Carron}, {Fortier}, {Pfyffer},
  {Benz}, {Mordasini}, \& {Swoboda}}]{Alibert2013}
{Alibert}, Y., {Carron}, F., {Fortier}, A., {et~al.} 2013, \aap, 558, A109

\bibitem[{{Almenara} {et~al.}(2018){Almenara}, {D{\'\i}az}, {Dorn}, {Bonfils},
  \& {Udry}}]{Almenara2018}
{Almenara}, J.~M., {D{\'\i}az}, R.~F., {Dorn}, C., {Bonfils}, X., \& {Udry}, S.
  2018, \mnras, 478, 460

\bibitem[{{Auclair-Desrotour} {et~al.}(2017){Auclair-Desrotour}, {Laskar},
  {Mathis}, \& {Correia}}]{Auclair2017}
{Auclair-Desrotour}, P., {Laskar}, J., {Mathis}, S., \& {Correia}, A.~C.~M.
  2017, \aap, 603, A108

\bibitem[{{Batygin} \& {Morbidelli}(2013)}]{BaMo2013}
{Batygin}, K. \& {Morbidelli}, A. 2013, Astron. Astrophys., 556, A28

\bibitem[{{Bean} {et~al.}(2021){Bean}, {Raymond}, \& {Owen}}]{Bean2021}
{Bean}, J.~L., {Raymond}, S.~N., \& {Owen}, J.~E. 2021, Journal of Geophysical
  Research (Planets), 126, e06639

\bibitem[{{Berger} {et~al.}(2020){Berger}, {Huber}, {van Saders}, {Gaidos},
  {Tayar}, \& {Kraus}}]{Berger2020}
{Berger}, T.~A., {Huber}, D., {van Saders}, J.~L., {et~al.} 2020, \aj, 159, 280

\bibitem[{{Bou{\'e}} {et~al.}(2012){Bou{\'e}}, {Oshagh}, {Montalto}, \&
  {Santos}}]{Boue2012}
{Bou{\'e}}, G., {Oshagh}, M., {Montalto}, M., \& {Santos}, N.~C. 2012, \mnras,
  422, L57

\bibitem[{{Carter} {et~al.}(2012){Carter}, {Agol}, {Chaplin}, {Basu},
  {Bedding}, {Buchhave}, {Christensen-Dalsgaard}, {Deck}, {Elsworth},
  {Fabrycky}, {Ford}, {Fortney}, {Hale}, {Handberg}, {Hekker}, {Holman},
  {Huber}, {Karoff}, {Kawaler}, {Kjeldsen}, {Lissauer}, {Lopez}, {Lund},
  {Lundkvist}, {Metcalfe}, {Miglio}, {Rogers}, {Stello}, {Borucki}, {Bryson},
  {Christiansen}, {Cochran}, {Geary}, {Gilliland}, {Haas}, {Hall}, {Howard},
  {Jenkins}, {Klaus}, {Koch}, {Latham}, {MacQueen}, {Sasselov}, {Steffen},
  {Twicken}, \& {Winn}}]{Carter2012}
{Carter}, J.~A., {Agol}, E., {Chaplin}, W.~J., {et~al.} 2012, Science, 337, 556

\bibitem[{{Coleman} {et~al.}(2019){Coleman}, {Leleu}, {Alibert}, \&
  {Benz}}]{Coleman2019}
{Coleman}, G.~A.~L., {Leleu}, A., {Alibert}, Y., \& {Benz}, W. 2019, \aap, 631,
  A7

\bibitem[{{Cubillos} {et~al.}(2017){Cubillos}, {Erkaev}, {Juvan}, {Fossati},
  {Johnstone}, {Lammer}, {Lendl}, {Odert}, \& {Kislyakova}}]{Cubillos2017}
{Cubillos}, P., {Erkaev}, N.~V., {Juvan}, I., {et~al.} 2017, \mnras, 466, 1868

\bibitem[{{Deck} {et~al.}(2014){Deck}, {Agol}, {Holman}, \&
  {Nesvorn{\'y}}}]{DeAgHo2014}
{Deck}, K.~M., {Agol}, E., {Holman}, M.~J., \& {Nesvorn{\'y}}, D. 2014, \apj,
  787, 132

\bibitem[{{Deck} {et~al.}(2013){Deck}, {Payne}, \& {Holman}}]{DePaHo2013}
{Deck}, K.~M., {Payne}, M., \& {Holman}, M.~J. 2013, \apj, 774, 129

\bibitem[{{Deline} {et~al.}(2022){Deline}, {Hooton}, {Lendl}, {Morris},
  {Salmon}, {Olofsson}, {Broeg}, {Ehrenreich}, {Beck}, {Brandeker}, {Hoyer},
  {Sulis}, {Van Grootel}, {Bourrier}, {Demangeon}, {Demory}, {Heng},
  {Parviainen}, {Serrano}, {Singh}, {Bonfanti}, {Fossati}, {Kitzmann}, {Sousa},
  {Wilson}, {Alibert}, {Alonso}, {Anglada}, {B{\'a}rczy}, {Barrado Navascues},
  {Barros}, {Baumjohann}, {Beck}, {Bekkelien}, {Benz}, {Billot}, {Bonfils},
  {Cabrera}, {Charnoz}, {Collier Cameron}, {Corral van Damme}, {Csizmadia},
  {Davies}, {Deleuil}, {Delrez}, {de Roche}, {Erikson}, {Fortier}, {Fridlund},
  {Futyan}, {Gandolfi}, {Gillon}, {G{\"u}del}, {Gutermann}, {Hasiba}, {Isaak},
  {Kiss}, {Laskar}, {Lecavelier des Etangs}, {Lovis}, {Magrin}, {Maxted},
  {Munari}, {Nascimbeni}, {Ottensamer}, {Pagano}, {Pall{\'e}}, {Peter},
  {Piotto}, {Pollacco}, {Queloz}, {Ragazzoni}, {Rando}, {Rauer}, {Ribas},
  {Santos}, {Scandariato}, {S{\'e}gransan}, {Simon}, {Smith}, {Steller},
  {Szab{\'o}}, {Thomas}, {Udry}, {Walter}, \& {Walton}}]{Deline2022}
{Deline}, A., {Hooton}, M.~J., {Lendl}, M., {et~al.} 2022, \aap, 659, A74

\bibitem[{{Delisle}(2017)}]{Delisle2017}
{Delisle}, J.~B. 2017, \aap, 605, A96

\bibitem[{{Delisle} {et~al.}(2017){Delisle}, {Correia}, {Leleu}, \&
  {Robutel}}]{DeCoLeRo2017}
{Delisle}, J.-B., {Correia}, A.~C.~M., {Leleu}, A., \& {Robutel}, P. 2017, aap,
  605, A37

\bibitem[{{Delisle} {et~al.}(2018){Delisle}, {S{\'e}gransan}, {Dumusque},
  {Diaz}, {Bouchy}, {Lovis}, {Pepe}, {Udry}, {Alonso}, {Benz}, {Coffinet},
  {Collier Cameron}, {Deleuil}, {Figueira}, {Gillon}, {Lo Curto}, {Mayor},
  {Mordasini}, {Motalebi}, {Moutou}, {Pollacco}, {Pompei}, {Queloz}, {Santos},
  \& {Wyttenbach}}]{Delisle2018}
{Delisle}, J.~B., {S{\'e}gransan}, D., {Dumusque}, X., {et~al.} 2018, \aap,
  614, A133

\bibitem[{{Dobrovolskis} \& {Borucki}(1996)}]{Dobrovolskis1996}
{Dobrovolskis}, A.~R. \& {Borucki}, W.~J. 1996, in \baas, Vol.~28, 1112

\bibitem[{{Dorn} {et~al.}(2017{\natexlab{a}}){Dorn}, {Venturini}, {Khan},
  {Heng}, {Alibert}, {Helled}, {Rivoldini}, \& {Benz}}]{Dorn2017}
{Dorn}, C., {Venturini}, J., {Khan}, A., {et~al.} 2017{\natexlab{a}}, 597, A37

\bibitem[{{Dorn} {et~al.}(2017{\natexlab{b}}){Dorn}, {Venturini}, {Khan},
  {Heng}, {Alibert}, {Helled}, {Rivoldini}, \& {Benz}}]{Dorn17}
{Dorn}, C., {Venturini}, J., {Khan}, A., {et~al.} 2017{\natexlab{b}}, 597, A37

\bibitem[{{Emsenhuber} {et~al.}(2021){Emsenhuber}, {Mordasini}, {Burn},
  {Alibert}, {Benz}, \& {Asphaug}}]{Emsenhuber2021}
{Emsenhuber}, A., {Mordasini}, C., {Burn}, R., {et~al.} 2021, \aap, 656, A69

\bibitem[{{Espinoza} \& {Jord{\'a}n}(2015)}]{EsJo2015}
{Espinoza}, N. \& {Jord{\'a}n}, A. 2015, \mnras, 450, 1879

\bibitem[{{Fulton} {et~al.}(2017){Fulton}, {Petigura}, {Howard}, {Isaacson},
  {Marcy}, {Cargile}, {Hebb}, {Weiss}, {Johnson}, {Morton}, {Sinukoff},
  {Crossfield}, \& {Hirsch}}]{Fulton2017}
{Fulton}, B.~J., {Petigura}, E.~A., {Howard}, A.~W., {et~al.} 2017, \aj, 154,
  109

\bibitem[{{Garc{\'\i}a-Melendo} \& {L{\'o}pez-Morales}(2011)}]{Garcia2011}
{Garc{\'\i}a-Melendo}, E. \& {L{\'o}pez-Morales}, M. 2011, \mnras, 417, L16

\bibitem[{{Go{\'z}dziewski} {et~al.}(2016){Go{\'z}dziewski}, {Migaszewski},
  {Panichi}, \& {Szuszkiewicz}}]{Gozdziewski2016}
{Go{\'z}dziewski}, K., {Migaszewski}, C., {Panichi}, F., \& {Szuszkiewicz}, E.
  2016, \mnras, 455, L104

\bibitem[{{Grimm} {et~al.}(2018){Grimm}, {Demory}, {Gillon}, {Dorn}, {Agol},
  {Burdanov}, {Delrez}, {Sestovic}, {Triaud}, {Turbet}, {Bolmont}, {Caldas},
  {Wit}, {Jehin}, {Leconte}, {Raymond}, {Grootel}, {Burgasser}, {Carey},
  {Fabrycky}, {Heng}, {Hernandez}, {Ingalls}, {Lederer}, {Selsis}, \&
  {Queloz}}]{Grimm2018}
{Grimm}, S.~L., {Demory}, B.-O., {Gillon}, M., {et~al.} 2018, \aap, 613, A68

\bibitem[{{Hadden} \& {Lithwick}(2016)}]{HaLi2016}
{Hadden}, S. \& {Lithwick}, Y. 2016, \apj, 828, 44

\bibitem[{{Hadden} \& {Lithwick}(2017)}]{HaLi2017}
{Hadden}, S. \& {Lithwick}, Y. 2017, \aj, 154, 5

\bibitem[{{Hakim} {et~al.}(2018){Hakim}, {Rivoldini}, {Van Hoolst},
  {Cottenier}, {Jaeken}, {Chust}, \& {Steinle-Neumann}}]{Hakim}
{Hakim}, K., {Rivoldini}, A., {Van Hoolst}, T., {et~al.} 2018, Icarus, 313, 61

\bibitem[{{Haldemann} {et~al.}(2020){Haldemann}, {Alibert}, {Mordasini}, \&
  {Benz}}]{Haldemann2020}
{Haldemann}, J., {Alibert}, Y., {Mordasini}, C., \& {Benz}, W. 2020, \aap, 643,
  A105

\bibitem[{{Henrard} \& {Lemaitre}(1983)}]{HenLe1983}
{Henrard}, J. \& {Lemaitre}, A. 1983, Celestial Mechanics, 30, 197

\bibitem[{{Holczer} {et~al.}(2016){Holczer}, {Mazeh}, {Nachmani},
  {Jontof-Hutter}, {Ford}, {Fabrycky}, {Ragozzine}, {Kane}, \&
  {Steffen}}]{Holczer2016}
{Holczer}, T., {Mazeh}, T., {Nachmani}, G., {et~al.} 2016, \apjs, 225, 9

\bibitem[{{Huber} {et~al.}(2013){Huber}, {Chaplin}, {Christensen-Dalsgaard},
  {Gilliland}, {Kjeldsen}, {Buchhave}, {Fischer}, {Lissauer}, {Rowe},
  {Sanchis-Ojeda}, {Basu}, {Handberg}, {Hekker}, {Howard}, {Isaacson},
  {Karoff}, {Latham}, {Lund}, {Lundkvist}, {Marcy}, {Miglio}, {Silva Aguirre},
  {Stello}, {Arentoft}, {Barclay}, {Bedding}, {Burke}, {Christiansen},
  {Elsworth}, {Haas}, {Kawaler}, {Metcalfe}, {Mullally}, \&
  {Thompson}}]{Huber2013}
{Huber}, D., {Chaplin}, W.~J., {Christensen-Dalsgaard}, J., {et~al.} 2013,
  \apj, 767, 127

\bibitem[{{Husser} {et~al.}(2013){Husser}, {Wende-von Berg}, {Dreizler},
  {Homeier}, {Reiners}, {Barman}, \& {Hauschildt}}]{Husser2013}
{Husser}, T.~O., {Wende-von Berg}, S., {Dreizler}, S., {et~al.} 2013, \aap,
  553, A6

\bibitem[{J{\'e}gou {et~al.}(2017)J{\'e}gou, Drozdzal, Vazquez, Romero, \&
  Bengio}]{jegou2017one}
J{\'e}gou, S., Drozdzal, M., Vazquez, D., Romero, A., \& Bengio, Y. 2017, in
  Proceedings of the IEEE conference on computer vision and pattern recognition
  workshops, 11--19

\bibitem[{{Jenkins} {et~al.}(2010){Jenkins}, {Caldwell}, {Chandrasekaran},
  {Twicken}, {Bryson}, {Quintana}, {Clarke}, {Li}, {Allen}, {Tenenbaum}, {Wu},
  {Klaus}, {Middour}, {Cote}, {McCauliff}, {Girouard}, {Gunter}, {Wohler},
  {Sommers}, {Hall}, {Uddin}, {Wu}, {Bhavsar}, {Van Cleve}, {Pletcher},
  {Dotson}, {Haas}, {Gilliland}, {Koch}, \& {Borucki}}]{Jenkins2010}
{Jenkins}, J.~M., {Caldwell}, D.~A., {Chandrasekaran}, H., {et~al.} 2010,
  \apjl, 713, L87

\bibitem[{{Jenkins} {et~al.}(2016){Jenkins}, {Twicken}, {McCauliff},
  {Campbell}, {Sanderfer}, {Lung}, {Mansouri-Samani}, {Girouard}, {Tenenbaum},
  {Klaus}, {Smith}, {Caldwell}, {Chacon}, {Henze}, {Heiges}, {Latham},
  {Morgan}, {Swade}, {Rinehart}, \& {Vanderspek}}]{Jenkins2016}
{Jenkins}, J.~M., {Twicken}, J.~D., {McCauliff}, S., {et~al.} 2016, in
  \procspie, Vol. 9913, Software and Cyberinfrastructure for Astronomy IV,
  99133E, tESS SPOC pipeline

\bibitem[{{Jontof-Hutter} {et~al.}(2016){Jontof-Hutter}, {Ford}, {Rowe},
  {Lissauer}, {Fabrycky}, {Van Laerhoven}, {Agol}, {Deck}, {Holczer}, \&
  {Mazeh}}]{Jontof2016}
{Jontof-Hutter}, D., {Ford}, E.~B., {Rowe}, J.~F., {et~al.} 2016, \apj, 820, 39

\bibitem[{{Jontof-Hutter} {et~al.}(2014){Jontof-Hutter}, {Lissauer}, {Rowe}, \&
  {Fabrycky}}]{Jontof-Hutter2014}
{Jontof-Hutter}, D., {Lissauer}, J.~J., {Rowe}, J.~F., \& {Fabrycky}, D.~C.
  2014, \apj, 785, 15

\bibitem[{{Kane} {et~al.}(2019){Kane}, {Ragozzine}, {Flowers}, {Holczer},
  {Mazeh}, \& {Relles}}]{Kane2019}
{Kane}, M., {Ragozzine}, D., {Flowers}, X., {et~al.} 2019, \aj, 157, 171

\bibitem[{{Kipping}(2013)}]{Kipping2013}
{Kipping}, D.~M. 2013, \mnras, 434, L51

\bibitem[{{Kov{\'a}cs} {et~al.}(2002){Kov{\'a}cs}, {Zucker}, \&
  {Mazeh}}]{Kovacs2002}
{Kov{\'a}cs}, G., {Zucker}, S., \& {Mazeh}, T. 2002, \aap, 391, 369

\bibitem[{{Kreidberg}(2015)}]{batman}
{Kreidberg}, L. 2015, \pasp, 127, 1161

\bibitem[{{Kurucz}(1979)}]{Kurucz1979}
{Kurucz}, R.~L. 1979, \apjs, 40, 1

\bibitem[{{Leconte} {et~al.}(2015){Leconte}, {Wu}, {Menou}, \&
  {Murray}}]{Leconte2015}
{Leconte}, J., {Wu}, H., {Menou}, K., \& {Murray}, N. 2015, Science, 347, 632

\bibitem[{{Lee} {et~al.}(2013){Lee}, {Fabrycky}, \& {Lin}}]{Lee2013}
{Lee}, M.~H., {Fabrycky}, D., \& {Lin}, D.~N.~C. 2013, \apj, 774, 52

\bibitem[{{Leleu} {et~al.}(2021{\natexlab{a}}){Leleu}, {Alibert}, {Hara},
  {Hooton}, {Wilson}, {Robutel}, {Delisle}, {Laskar}, {Hoyer}, {Lovis},
  {Bryant}, {Ducrot}, {Cabrera}, {Delrez}, {Acton}, {Adibekyan}, {Allart},
  {Allende Prieto}, {Alonso}, {Alves}, {Anderson}, {Angerhausen}, {Anglada
  Escud{\'e}}, {Asquier}, {Barrado}, {Barros}, {Baumjohann}, {Bayliss}, {Beck},
  {Beck}, {Bekkelien}, {Benz}, {Billot}, {Bonfanti}, {Bonfils}, {Bouchy},
  {Bourrier}, {Bou{\'e}}, {Brandeker}, {Broeg}, {Buder}, {Burdanov},
  {Burleigh}, {B{\'a}rczy}, {Cameron}, {Chamberlain}, {Charnoz}, {Cooke},
  {Corral Van Damme}, {Correia}, {Cristiani}, {Damasso}, {Davies}, {Deleuil},
  {Demangeon}, {Demory}, {Di Marcantonio}, {Di Persio}, {Dumusque},
  {Ehrenreich}, {Erikson}, {Figueira}, {Fortier}, {Fossati}, {Fridlund},
  {Futyan}, {Gandolfi}, {Garc{\'\i}a Mu{\~n}oz}, {Garcia}, {Gill}, {Gillen},
  {Gillon}, {Goad}, {Gonz{\'a}lez Hern{\'a}ndez}, {Guedel}, {G{\"u}nther},
  {Haldemann}, {Henderson}, {Heng}, {Hogan}, {Isaak}, {Jehin}, {Jenkins},
  {Jord{\'a}n}, {Kiss}, {Kristiansen}, {Lam}, {Lavie}, {Lecavelier des Etangs},
  {Lendl}, {Lillo-Box}, {Lo Curto}, {Magrin}, {Martins}, {Maxted}, {McCormac},
  {Mehner}, {Micela}, {Molaro}, {Moyano}, {Murray}, {Nascimbeni}, {Nunes},
  {Olofsson}, {Osborn}, {Oshagh}, {Ottensamer}, {Pagano}, {Pall{\'e}},
  {Pedersen}, {Pepe}, {Persson}, {Peter}, {Piotto}, {Polenta}, {Pollacco},
  {Poretti}, {Pozuelos}, {Queloz}, {Ragazzoni}, {Rando}, {Ratti}, {Rauer},
  {Raynard}, {Rebolo}, {Reimers}, {Ribas}, {Santos}, {Scandariato},
  {Schneider}, {Sebastian}, {Sestovic}, {Simon}, {Smith}, {Sousa}, {Sozzetti},
  {Steller}, {Su{\'a}rez Mascare{\~n}o}, {Szab{\'o}}, {S{\'e}gransan},
  {Thomas}, {Thompson}, {Tilbrook}, {Triaud}, {Turner}, {Udry}, {Van Grootel},
  {Venus}, {Verrecchia}, {Vines}, {Walton}, {West}, {Wheatley}, {Wolter}, \&
  {Zapatero Osorio}}]{Leleu2021}
{Leleu}, A., {Alibert}, Y., {Hara}, N.~C., {et~al.} 2021{\natexlab{a}}, \aap,
  649, A26

\bibitem[{{Leleu} {et~al.}(2021{\natexlab{b}}){Leleu}, {Chatel}, {Udry},
  {Alibert}, {Delisle}, \& {Mardling}}]{RIVERS1}
{Leleu}, A., {Chatel}, G., {Udry}, S., {et~al.} 2021{\natexlab{b}}, \aap, 655,
  A66

\bibitem[{{Leleu} {et~al.}(2022){Leleu}, {Delisle}, {Mardling}, {Udry},
  {Chatel}, {Alibert}, \& {Eggenberger}}]{RIVERS2}
{Leleu}, A., {Delisle}, J.~B., {Mardling}, R., {et~al.} 2022, arXiv e-prints,
  arXiv:2201.11459

\bibitem[{{Libby-Roberts} {et~al.}(2020){Libby-Roberts}, {Berta-Thompson},
  {D{\'e}sert}, {Masuda}, {Morley}, {Lopez}, {Deck}, {Fabrycky}, {Fortney},
  {Line}, {Sanchis-Ojeda}, \& {Winn}}]{Libby-Roberts2020}
{Libby-Roberts}, J.~E., {Berta-Thompson}, Z.~K., {D{\'e}sert}, J.-M., {et~al.}
  2020, \aj, 159, 57

\bibitem[{{Lithwick} {et~al.}(2012){Lithwick}, {Xie}, \& {Wu}}]{Lithwick2012}
{Lithwick}, Y., {Xie}, J., \& {Wu}, Y. 2012, \apj, 761, 122

\bibitem[{{Lopez} \& {Fortney}(2014)}]{LopezFortney14}
{Lopez}, E.~D. \& {Fortney}, J.~J. 2014, 792, 1

\bibitem[{{Luger} {et~al.}(2017){Luger}, {Foreman-Mackey}, \&
  {Hogg}}]{Luger2017linear}
{Luger}, R., {Foreman-Mackey}, D., \& {Hogg}, D.~W. 2017, Research Notes of the
  American Astronomical Society, 1, 7

\bibitem[{{Marboeuf} {et~al.}(2014){Marboeuf}, {Thiabaud}, {Alibert}, {Cabral},
  \& {Benz}}]{Marboeuf}
{Marboeuf}, U., {Thiabaud}, A., {Alibert}, Y., {Cabral}, N., \& {Benz}, W.
  2014, 570, A36

\bibitem[{{Masuda}(2014)}]{Masuda2014}
{Masuda}, K. 2014, \apj, 783, 53

\bibitem[{{Mazeh} {et~al.}(2013){Mazeh}, {Nachmani}, {Holczer}, {Fabrycky},
  {Ford}, {Sanchis-Ojeda}, {Sokol}, {Rowe}, {Zucker}, {Agol}, {Carter},
  {Lissauer}, {Quintana}, {Ragozzine}, {Steffen}, \& {Welsh}}]{Mazeh2013}
{Mazeh}, T., {Nachmani}, G., {Holczer}, T., {et~al.} 2013, \apjs, 208, 16

\bibitem[{{Mills} \& {Fabrycky}(2017)}]{Mills2017}
{Mills}, S.~M. \& {Fabrycky}, D.~C. 2017, \apjl, 838, L11

\bibitem[{{Mills} {et~al.}(2016){Mills}, {Fabrycky}, {Migaszewski}, {Ford},
  {Petigura}, \& {Isaacson}}]{Mills16}
{Mills}, S.~M., {Fabrycky}, D.~C., {Migaszewski}, C., {et~al.} 2016, \nat, 533,
  509

\bibitem[{{Mills} \& {Mazeh}(2017)}]{MillsMazeh2017}
{Mills}, S.~M. \& {Mazeh}, T. 2017, \apjl, 839, L8

\bibitem[{{Mordasini}(2018)}]{Mordasini2018}
{Mordasini}, C. 2018, {Planetary Population Synthesis}, 143

\bibitem[{{Mordasini} {et~al.}(2009){Mordasini}, {Alibert}, {Benz}, \&
  {Naef}}]{Mordasini2009}
{Mordasini}, C., {Alibert}, Y., {Benz}, W., \& {Naef}, D. 2009, \aap, 501, 1161

\bibitem[{{Nesvorny} {et~al.}(2021){Nesvorny}, {Chrenko}, \&
  {Flock}}]{Nesvorny2021}
{Nesvorny}, D., {Chrenko}, O., \& {Flock}, M. 2021, arXiv e-prints,
  arXiv:2110.09577

\bibitem[{{Nesvorn{\'y}} {et~al.}(2013){Nesvorn{\'y}}, {Kipping}, {Terrell},
  {Hartman}, {Bakos}, \& {Buchhave}}]{Nesvorny2013}
{Nesvorn{\'y}}, D., {Kipping}, D., {Terrell}, D., {et~al.} 2013, \apj, 777, 3

\bibitem[{{Nesvorn{\'y}} \& {Vokrouhlick{\'y}}(2014)}]{NeVo2014}
{Nesvorn{\'y}}, D. \& {Vokrouhlick{\'y}}, D. 2014, apj, 790, 58

\bibitem[{{Nesvorn{\'y}} \& {Vokrouhlick{\'y}}(2016)}]{NeVo2016}
{Nesvorn{\'y}}, D. \& {Vokrouhlick{\'y}}, D. 2016, \apj, 823, 72

\bibitem[{{Otegi} {et~al.}(2020){Otegi}, {Bouchy}, \& {Helled}}]{Otegi2020}
{Otegi}, J.~F., {Bouchy}, F., \& {Helled}, R. 2020, \aap, 634, A43

\bibitem[{{Panichi} {et~al.}(2019){Panichi}, {Migaszewski}, \&
  {Go{\'z}dziewski}}]{Panichi2019}
{Panichi}, F., {Migaszewski}, C., \& {Go{\'z}dziewski}, K. 2019, \mnras, 485,
  4601

\bibitem[{{Ragozzine} \& {Holman}(2010)}]{RaHo2010}
{Ragozzine}, D. \& {Holman}, M.~J. 2010, arXiv e-prints, arXiv:1006.3727

\bibitem[{{Rowe} {et~al.}(2014){Rowe}, {Bryson}, {Marcy}, {Lissauer},
  {Jontof-Hutter}, {Mullally}, {Gilliland}, {Issacson}, {Ford}, {Howell},
  {Borucki}, {Haas}, {Huber}, {Steffen}, {Thompson}, {Quintana}, {Barclay},
  {Still}, {Fortney}, {Gautier}, {Hunter}, {Caldwell}, {Ciardi}, {Devore},
  {Cochran}, {Jenkins}, {Agol}, {Carter}, \& {Geary}}]{Rowe2014}
{Rowe}, J.~F., {Bryson}, S.~T., {Marcy}, G.~W., {et~al.} 2014, \apj, 784, 45

\bibitem[{{Rowe} {et~al.}(2015){Rowe}, {Coughlin}, {Antoci}, {Barclay},
  {Batalha}, {Borucki}, {Burke}, {Bryson}, {Caldwell}, {Campbell},
  {Catanzarite}, {Christiansen}, {Cochran}, {Gilliland}, {Girouard}, {Haas},
  {He{\l}miniak}, {Henze}, {Hoffman}, {Howell}, {Huber}, {Hunter},
  {Jang-Condell}, {Jenkins}, {Klaus}, {Latham}, {Li}, {Lissauer}, {McCauliff},
  {Morris}, {Mullally}, {Ofir}, {Quarles}, {Quintana}, {Sabale}, {Seader},
  {Shporer}, {Smith}, {Steffen}, {Still}, {Tenenbaum}, {Thompson}, {Twicken},
  {Van Laerhoven}, {Wolfgang}, \& {Zamudio}}]{Rowe2015}
{Rowe}, J.~F., {Coughlin}, J.~L., {Antoci}, V., {et~al.} 2015, \apjs, 217, 16

\bibitem[{{Rowe} \& {Thompson}(2015)}]{RoTho2015}
{Rowe}, J.~F. \& {Thompson}, S.~E. 2015, arXiv e-prints, arXiv:1504.00707

\bibitem[{{Sotin} {et~al.}(2007){Sotin}, {Grasset}, \& {Mocquet}}]{Sotin}
{Sotin}, C., {Grasset}, O., \& {Mocquet}, A. 2007, Icarus, 191, 337

\bibitem[{{Thiabaud} {et~al.}(2014){Thiabaud}, {Marboeuf}, {Alibert}, {Cabral},
  {Leya}, \& {Mezger}}]{Thiabaud}
{Thiabaud}, A., {Marboeuf}, U., {Alibert}, Y., {et~al.} 2014, 562, A27

\bibitem[{{Thiabaud} {et~al.}(2015){Thiabaud}, {Marboeuf}, {Alibert}, {Leya},
  \& {Mezger}}]{Thiabaud2015}
{Thiabaud}, A., {Marboeuf}, U., {Alibert}, Y., {Leya}, I., \& {Mezger}, K.
  2015, \aap, 580, A30

\bibitem[{{Turbet} {et~al.}(2020){Turbet}, {Bolmont}, {Ehrenreich}, {Gratier},
  {Leconte}, {Selsis}, {Hara}, \& {Lovis}}]{Turbet2020}
{Turbet}, M., {Bolmont}, E., {Ehrenreich}, D., {et~al.} 2020, \aap, 638, A41

\bibitem[{{Vissapragada} {et~al.}(2020){Vissapragada}, {Jontof-Hutter},
  {Shporer}, {Knutson}, {Liu}, {Thorngren}, {Lee}, {Chachan}, {Mawet},
  {Millar-Blanchaer}, {Nilsson}, {Tinyanont}, {Vasisht}, \&
  {Wright}}]{Vissapragada2020}
{Vissapragada}, S., {Jontof-Hutter}, D., {Shporer}, A., {et~al.} 2020, \aj,
  159, 108

\bibitem[{{Weiss} \& {Marcy}(2014)}]{WeissMarcy2014}
{Weiss}, L.~M. \& {Marcy}, G.~W. 2014, \apjl, 783, L6

\bibitem[{{Wu} \& {Lithwick}(2013)}]{WuLi2013}
{Wu}, Y. \& {Lithwick}, Y. 2013, \apj, 772, 74

\bibitem[{{Xie} {et~al.}(2014){Xie}, {Wu}, \& {Lithwick}}]{Xie2014}
{Xie}, J.-W., {Wu}, Y., \& {Lithwick}, Y. 2014, \apj, 789, 165

\bibitem[{{Zeng} {et~al.}(2019){Zeng}, {Jacobsen}, {Sasselov}, {Petaev},
  {Vanderburg}, {Lopez-Morales}, {Perez-Mercader}, {Mattsson}, {Li}, {Heising},
  {Bonomo}, {Damasso}, {Berger}, {Cao}, {Levi}, \& {Wordsworth}}]{Zeng19}
{Zeng}, L., {Jacobsen}, S.~B., {Sasselov}, D.~D., {et~al.} 2019, Proceedings of
  the National Academy of Science, 116, 9723

\bibitem[{{Zeng} {et~al.}(2016){Zeng}, {Sasselov}, \& {Jacobsen}}]{Zeng2016}
{Zeng}, L., {Sasselov}, D.~D., \& {Jacobsen}, S.~B. 2016, \apj, 819, 127

\bibitem[{{Zhu} {et~al.}(2018){Zhu}, {Petrovich}, {Wu}, {Dong}, \&
  {Xie}}]{Zhu2018}
{Zhu}, W., {Petrovich}, C., {Wu}, Y., {Dong}, S., \& {Xie}, J. 2018, \apj, 860,
  101

\end{thebibliography}

\appendix

\section{Eccentricity and longitude of periastron posterior shapes}
\label{ap:post_shape}

\begin{figure*}[!ht]
\begin{center}
\includegraphics[width=0.49\textwidth]{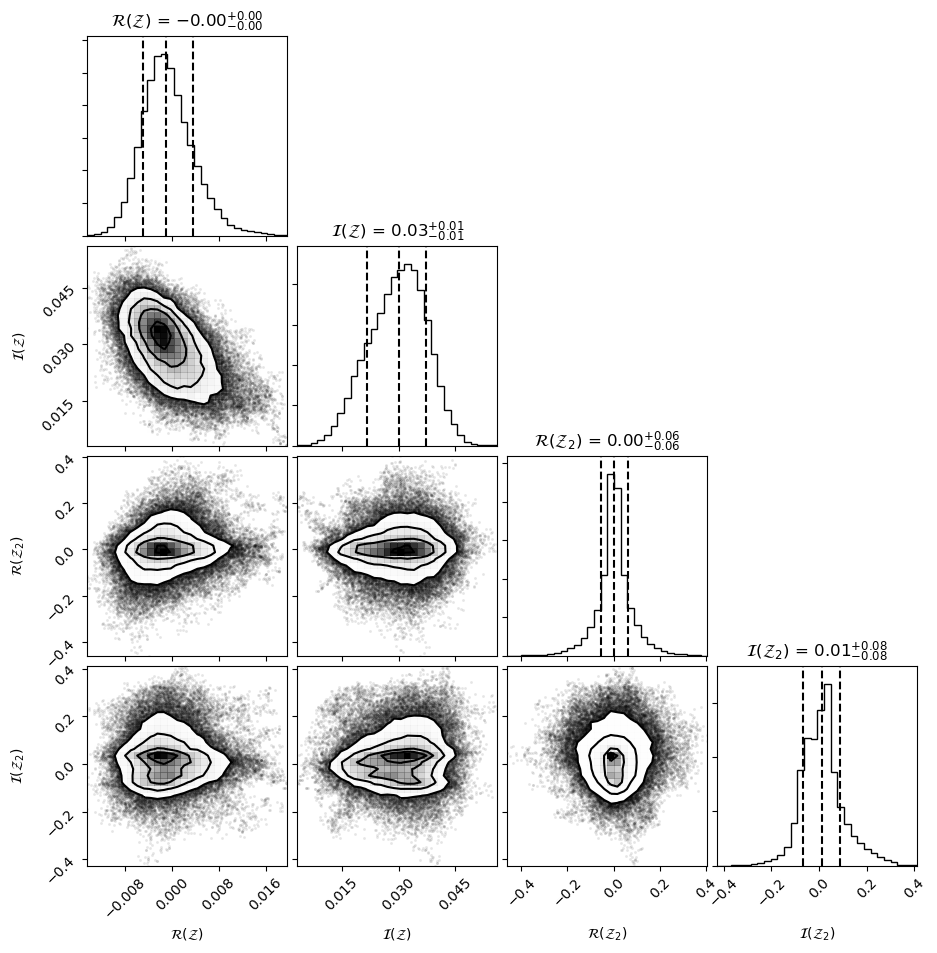}
\includegraphics[width=0.49\textwidth]{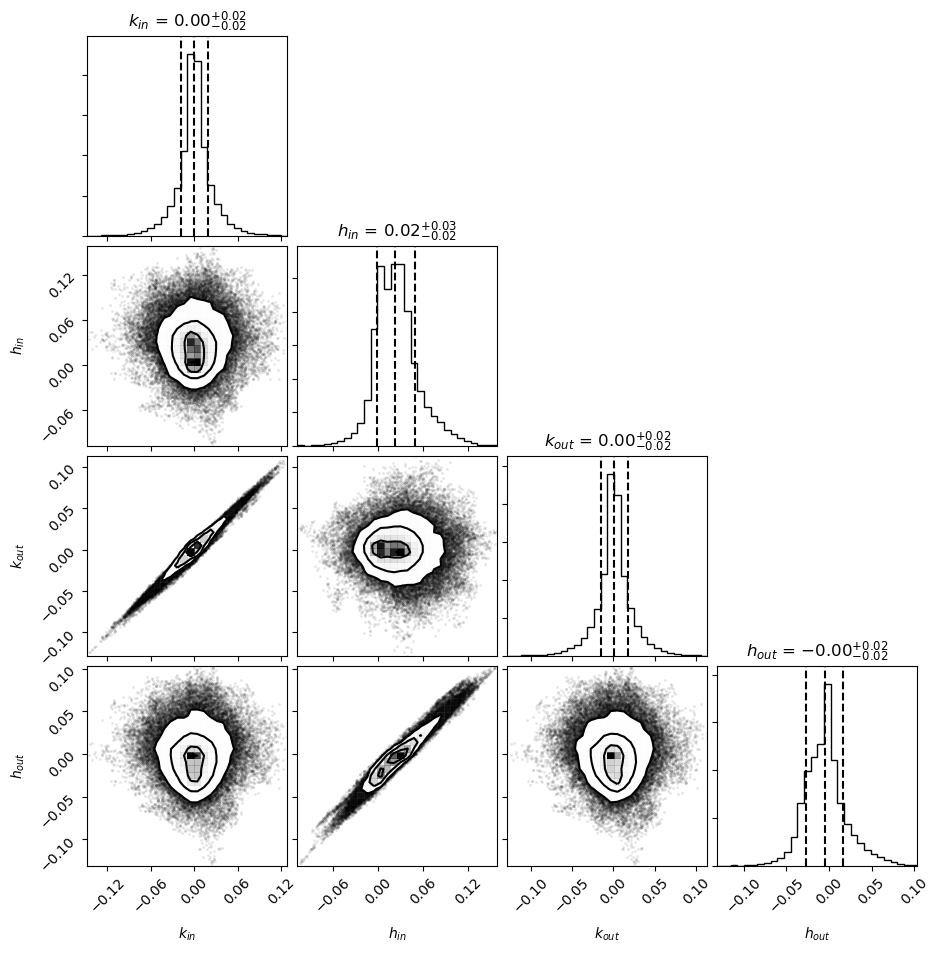}\\
\includegraphics[width=0.49\textwidth]{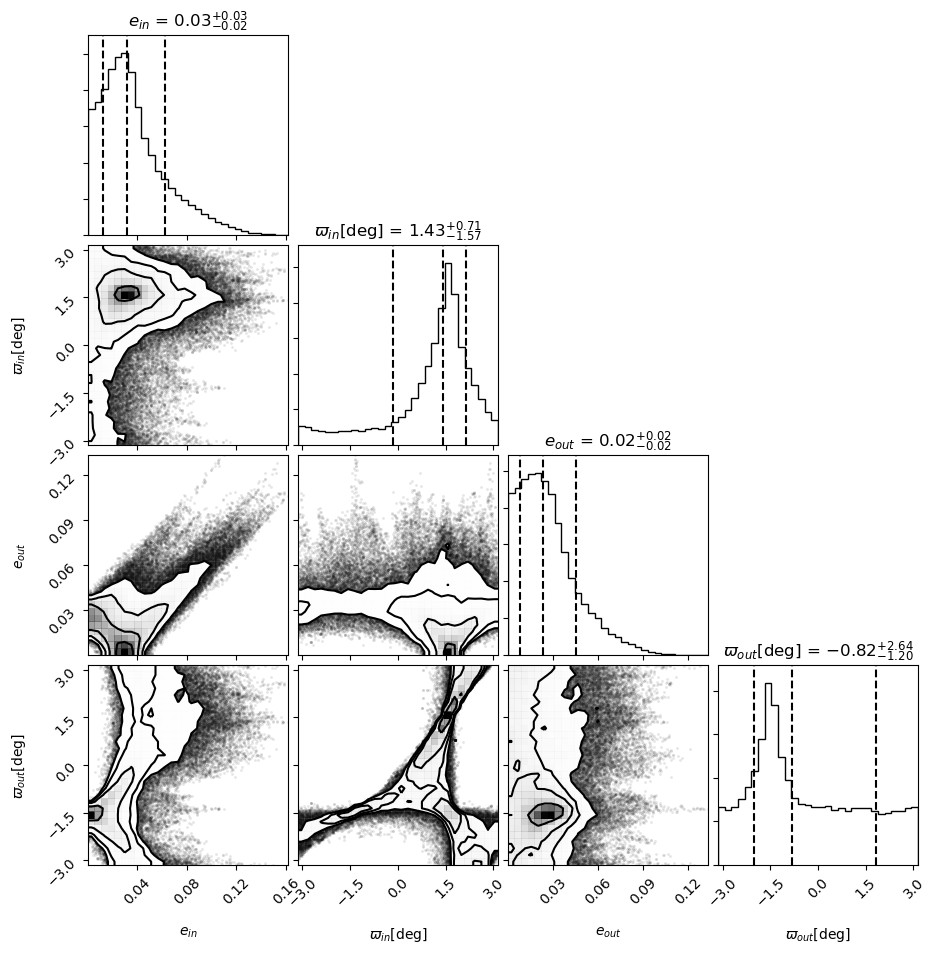}
\includegraphics[width=0.49\textwidth]{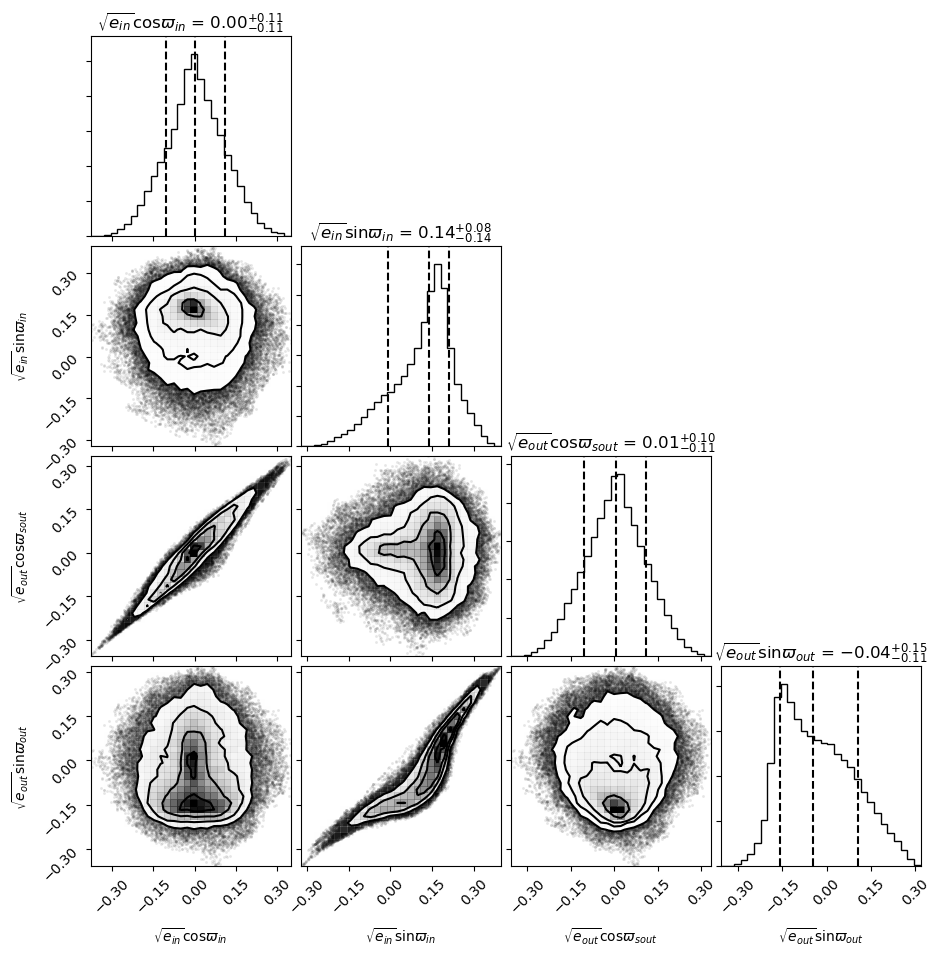}
\caption{\label{fig:post_ZZ2} Corner plots of the eccentricities and longitudes of periastron of the \default posterior of Kepler-345 b and c. Top left corner shows the real and imaginary part of $\cZ$ and $\cZ_2$, top right corner shows the $k_i=e_i\cos \varpi_i$ and $h_i=e_i\sin \varpi_i$ variables, bottom left corner shows the $e_i$ and $\varpi_i$ variables, and the bottom right corner shows the $\sqrt{e_i} \cos \varpi_i$ and $\sqrt{e_i} \sin \varpi_i$ variables. }
\end{center}
\end{figure*}

Figure \ref{fig:post_ZZ2} shows the correlation between parameters for various choices of coordinates.

\section{Analytical modeling of TTVs}
\label{ap:anaTTVs}

{Table \ref{tab:anaTTVs} shows the peak-to-peak amplitude of the sinusoidal approximation of the best fit of the photo-dynamical model (section \ref{sec:amplitude}). The table also shows the different TTVs contributions from the analytical model of \cite{HaLi2016}, valid outside first- or second-order MMRs. These coefficients are computed as follows: we randomly select 400 samples of the \textit{final} posterior. For each of these initial conditions, we compute the analytical TTVs for each subsequent pair of planets. We compute the fundamental, secondary, and chopping signal separately based on appendix B of \cite{HaLi2016}, see section \ref{sec:ZZ2} for a discussion on these terms. For the chopping signal, we compute the first 10 terms of the series, see eq. (39) and (54) of \cite[][]{HaLi2016}. Then, for the fundamental, secondary, and the sum of the chopping terms, we compute the amplitude by subtracting their minimum value from their maximum value over the duration of the Kepler mission. This results in 400 estimations of amplitude for each term. In the table, we display the median and .16 and .84 quantiles error for each amplitude across these 400 samples.} 

{We did not display the results for Kepler-60, since the planets are inside the MMRs and therefore the model is not valid. We also note that the model does not work properly for Kepler-57, which might be strongly affected by the proximity of the resonance or require the consideration of additional terms in eccentricity. For all other pairs, the amplitude of the sinusoidal approximation is similar to the estimated amplitude of the fundamental harmonic. The table gives an idea of the relative size of the secondary and chopping terms that need to be constrained in order to break the mass/eccentricity degeneracy. However, the interpretation of these results is not straightforward. Firstly, the difficulty to detect a harmonic depend not only on its amplitude but also strongly on the SNR$_i$ of the planet and the availability of short cadence data. Secondly, the analytical TTV model was not directly fit to the data: small errors on secondary and chopping signals do not necessarily imply that we were able to measure these contributions with high precision, but that the overall constraints we got on the system allowed to give a precise estimation of what should be the amplitude of these terms. For example, the small uncertainties on the chopping terms of Kepler-57 illustrate this point. Thirdly, the model is only considering a pair of planets. }

\begin{table*} 
\begin{small} 
\caption{TTV amplitude of 19 Kepler pairs} 
\label{tab:MR} 
\centering 
\begin{tabular}{llllllllll} 
\hline 
Pair & $P_{out}/P_{in}$ & $Amp_{in}$ & $Amp_{out}$ & $\sigma_{F,in}$ & $\sigma_{F,out}$ & $\sigma_{S,in}$ & $\sigma_{S,out}$  & $\sigma_{C,in}$ & $\sigma_{C,out}$ \\ 
\hline\hline 
Kepler-23 bc & $1.51$&$97.13$&$45.84$ & $95.10_{-3.97}^{+4.43}$ & $41.55_{-1.07}^{+1.22}$ & $4.82_{-0.52}^{+0.47}$ & $2.12_{-0.18}^{+0.22}$ & $2.45_{-0.22}^{+0.24}$ & $0.869_{-0.078}^{+0.090}$ \\ 
Kepler-23 cd & $1.42$&$45.84$&$6.11$ & $-$ & $-$ & $0.41_{-0.18}^{+0.33}$ & $0.99_{-0.41}^{+0.68}$ & $4.81_{-0.93}^{+0.95}$ & $10.73_{-0.83}^{+1.05}$ \\ 
Kepler-24 bc & $1.52$&$62.66$&$111.83$ & $74.69_{-3.46}^{+3.69}$ & $114.62_{-5.49}^{+6.27}$ & $3.05_{-0.21}^{+0.18}$ & $4.82_{-0.28}^{+0.33}$ & $3.55_{-0.36}^{+0.51}$ & $4.32_{-0.48}^{+0.61}$ \\ 
Kepler-24 ce & $1.54$&$111.83$&$39.97$ & $0.70_{-0.34}^{+1.31}$ & $45.02_{-7.52}^{+7.93}$ & $0.06_{-0.03}^{+0.12}$ & $4.20_{-1.37}^{+1.45}$ & $0.12_{-0.06}^{+0.22}$ & $6.04_{-0.75}^{+0.78}$ \\ 
Kepler-26 bc & $1.41$&$21.69$&$16.66$ & $-$ & $-$ & $18.73_{-0.81}^{+0.75}$ & $14.86_{-0.84}^{+0.82}$ & $15.76_{-0.65}^{+0.60}$ & $13.18_{-0.83}^{+0.75}$ \\ 
Kepler-49 bc & $1.52$&$20.17$&$36.80$ & $9.18_{-2.34}^{+3.96}$ & $15.79_{-1.97}^{+4.75}$ & $0.228_{-0.035}^{+0.046}$ & $0.482_{-0.073}^{+0.088}$ & $4.72_{-0.36}^{+0.36}$ & $5.96_{-0.38}^{+0.38}$ \\ 
Kepler-28 bc & $1.52$&$19.53$&$20.72$ & $18.25_{-1.02}^{+0.92}$ & $19.56_{-1.00}^{+1.10}$ & $1.45_{-0.25}^{+0.31}$ & $1.59_{-0.27}^{+0.34}$ & $0.80_{-0.14}^{+0.22}$ & $0.70_{-0.11}^{+0.20}$ \\ 
Kepler-52 bc & $2.08$&$36.30$&$17.96$ & $33.32_{-3.29}^{+2.59}$ & $15.63_{-2.43}^{+1.88}$ & $2.53_{-0.56}^{+0.67}$ & $1.08_{-0.26}^{+0.35}$ & $0.80_{-0.13}^{+0.16}$ & $0.105_{-0.018}^{+0.024}$ \\ 
Kepler-54 bc & $1.51$&$127.90$&$255.26$ & $140.49_{-2.78}^{+2.55}$ & $278.65_{-3.97}^{+4.63}$ & $4.77_{-0.33}^{+0.50}$ & $9.67_{-0.63}^{+0.95}$ & $1.53_{-0.11}^{+0.10}$ & $2.43_{-0.18}^{+0.15}$ \\ 
Kepler-57 bc & $2.03$&$10.92$&$65.86$ & $60.33_{-6.10}^{+5.19}$ & $338.5_{-23.7}^{+27.4}$ & $3.73_{-0.36}^{+0.31}$ & $20.21_{-1.45}^{+1.39}$ & $0.245_{-0.035}^{+0.036}$ & $0.415_{-0.057}^{+0.061}$ \\ 
Kepler-58 bc & $1.52$&$37.30$&$77.49$ & $30.48_{-5.61}^{+5.56}$ & $69.69_{-7.25}^{+9.43}$ & $2.85_{-0.71}^{+1.03}$ & $7.06_{-1.92}^{+2.11}$ & $1.59_{-0.45}^{+0.99}$ & $3.12_{-0.82}^{+1.39}$ \\ 
Kepler-85 bc & $1.51$&$83.20$&$101.12$ & $81.29_{-4.48}^{+5.31}$ & $93.34_{-4.13}^{+5.19}$ & $3.34_{-0.37}^{+0.52}$ & $3.92_{-0.43}^{+0.63}$ & $0.93_{-0.17}^{+0.21}$ & $0.85_{-0.15}^{+0.19}$ \\ 
Kepler-128 bc & $1.51$&$163.99$&$230.13$ & $139.24_{-3.98}^{+3.36}$ & $207.22_{-5.69}^{+6.59}$ & $13.47_{-1.27}^{+1.50}$ & $20.17_{-1.66}^{+2.52}$ & $1.82_{-0.23}^{+0.21}$ & $2.20_{-0.27}^{+0.28}$ \\ 
Kepler-176 cd & $2.02$&$45.47$&$47.32$ & $56.0_{-12.3}^{+12.7}$ & $64.2_{-14.4}^{+18.9}$ & $3.31_{-1.32}^{+1.40}$ & $3.84_{-1.40}^{+1.61}$ & $0.107_{-0.021}^{+0.049}$ & $0.036_{-0.008}^{+0.020} $\\ 
Kepler-305 bc & $1.51$&$10.46$&$28.32$ & $10.89_{-2.68}^{+2.58}$ & $22.38_{-4.40}^{+3.80}$ & $0.140_{-0.038}^{+0.065}$ & $0.30_{-0.06}^{+0.14}$ & $1.06_{-0.16}^{+0.26}$ & $1.67_{-0.27}^{+0.41}$ \\ 
Kepler-305 cd & $2.02$&$28.32$&$8.54$ & $21.1_{-8.3}^{+19.3}$ & $20.6_{-8.0}^{+11.9}$ & $0.17_{-0.08}^{+0.43}$ & $0.16_{-0.06}^{+0.27}$ & $0.345_{-0.057}^{+0.085}$ & $0.091_{-0.015}^{+0.019}$ \\ 
Kepler-345 bc & $1.27$&$14.52$&$9.21$ & $13.66_{-3.85}^{+4.81}$ & $8.32_{-1.40}^{+1.47}$ & $0.82_{-0.29}^{+0.50}$ & $0.53_{-0.16}^{+0.17}$ & $6.70_{-1.90}^{+2.24}$ & $3.85_{-0.73}^{+0.95}$ \\ 
\hline 
\end{tabular} 
\tablefoot{  
\label{tab:anaTTVs} Peak-to-peak amplitude of the sinusoidal approximation for the inner and outer planet of the pair, and the peak-to-peak amplitude of the fundamental ($\sigma_{F}$), secondary ($\sigma_{S}$) and chopping signal ($\sigma_{C}$) along the duration of the Kepler mission, from the analytical model of planets outside of first and second order MMRs by \cite{HaLi2016}. All amplitudes are in minutes.}
\end{small} 
\end{table*}

\section{B-splines and marginalization of the likelihood}
\label{sec:lmarg}

In this appendix, we describe the B-spline model
used to account for stellar variations and instrumental systematics,
as well as the method we use to efficiently compute the likelihood
marginalized over the B-spline parameters.
A cubic B-spline is a piecewise third order polynomial that is twice continuously differentiable
everywhere.
We assume here regularly spaced knots and denote by $\tau$ the time lag between two knots.
The value of $\tau$ is chosen
in order to avoid over-fitting short term variations associated with the transits
but still model as much as possible of stellar variations and instrumental systematics.
The procedure used to select the value of $\tau$ is described in Sec.~\ref{sec:preprocess}.
For a time series with large interruptions ($\Delta t > 4 \tau$),
the B-splines over each of the segments of continuous observations
are independent from each other.
Thus, the full marginal likelihood is simply the product
of the marginal likelihoods over each segment.
We thus consider here a single segment with continuous observations ($\Delta t < 4 \tau$).
For a segment with time span $T$, we have $N = \lceil T/\tau \rceil$
pieces ($N+1$ knots).
We center the time series in the sense that
we set the positions ($\tau_k$) of the knots such that
the lag between the first knot and the first measurement ($t_1-\tau_1$)
is the same as the lag between the last measurement and the last knot ($\tau_{N+1}-t_n$).
The B-spline is defined as a linear combination of $N+3$ splines.
Each piece is modeled as a combination of four splines
and, reciprocally, each spline is defined over four consecutive pieces (except on the edges).
The parameters $\eta$ of the model are the $N+3$ coefficients
appearing in the linear combination of the splines.
For $t\in[\tau_k, \tau_{k+1}]$,
we have
\begin{align}
  b(\eta, t) = & \ \eta_k \left(1-\delta\right)^3
  + \eta_{k+1} \left(3\delta^3-6\delta^2+4\right)\nonumber                \\
               & + \eta_{k+2} \left(-3\delta^3+3\delta^2+3\delta+1\right)
  + \eta_{k+3} \delta^3,
\end{align}
where $\delta = t - \tau_k$.
The time series $b(\eta, t)$ is thus of the form
\begin{equation}
  b(\eta, t) = B \eta
\end{equation}
where $B$ is the $(n\times (N+3))$ matrix defined as
\begin{equation}
  \setlength{\arraycolsep}{1pt}
  B = \begin{pmatrix}
    \beta_{1,1}   & \dots           & \beta_{1,4}   & 0                 & \dots               & \dots             & 0                   \\
    \vdots        & \vdots          & \vdots        & \vdots            & \vdots              & \vdots            & \vdots              \\
    \beta_{n_1,1} & \dots           & \beta_{n_1,4} & 0                 & \dots               & \dots             & 0                   \\
    0             & \beta_{n_1+1,1} & \dots         & \beta_{n_1+1,4}   & 0                   & \dots             & 0                   \\
    \vdots        & \vdots          & \vdots        & \vdots            & \vdots              & \vdots            & \vdots              \\
    0             & \beta_{n_2,1}   & \dots         & \beta_{n_2,4}     & 0                   & \dots             & 0                   \\
    \vdots        & \ddots          & \ddots        & \ddots            & \ddots              & \ddots            & \vdots              \\
    0             & \dots           & 0             & \beta_{n_{N-1},1} & \dots               & \beta_{n_{N-1},4} & 0                   \\
    0             & \dots           & \dots         & 0                 & \beta_{n_{N-1}+1,1} & \dots             & \beta_{n_{N-1}+1,4} \\
    \vdots        & \vdots          & \vdots        & \vdots            & \vdots              & \vdots            & \vdots              \\
    0             & \dots           & \dots         & 0                 & \beta_{n,1}         & \dots             & \beta_{n,4}
  \end{pmatrix},
\end{equation}
with $n_k$ the index of the last measurement that lies in the range $[\tau_k, \tau_{k+1}]$
and
\begin{align}
  \beta_{i,1} & = \left(1-\delta_i\right)^3,\nonumber            \\
  \beta_{i,2} & = 3\delta_i^3-6\delta_i^2+4,\nonumber            \\
  \beta_{i,3} & = -3\delta_i^3+3\delta_i^2+3\delta_i+1,\nonumber \\
  \beta_{i,4} & = \delta_i^3.
\end{align}
The matrix $B$ can thus be stored in an efficient manner
in the form of the $(n\times 4)$ matrix $\beta$.

We now aim at computing the marginal likelihood of Eq.~(\ref{eq:lmarg}).
For this purpose, we first need to set a prior on the parameters $\eta$.
We assume for $\eta$ a centered Gaussian prior with covariance $\Lambda$,
which is independent of the other parameters ($\theta,\sigma_\mathrm{jit.}$):
\begin{equation}
  p(\eta|\theta,\sigma_\mathrm{jit.}) = p(\eta) = \frac{1}{\sqrt{|2\pi\Lambda|}} \exp\left(-\frac{1}{2} \eta\t \Lambda^{-1} \eta \right).
\end{equation}
We additionally assume $\Lambda$ to be diagonal in the following.
For a given set of parameters $\theta$, we can compute the transit model $m(\theta, t)$,
and define
\begin{equation}
  A(\theta, t) = m(\theta,t) * B(t),
\end{equation}
where $*$ denotes the Hadamard (element-wise) product of each column of $B$ by the vector $m$.
The matrix $A$ possesses the exact same structure as $B$
and can be stored using the $n\times 4$ matrix $\alpha = m * \beta$.
The marginal likelihood of Eq.~(\ref{eq:lmarg}) can then be rewritten as
\begin{align}
  \mathcal{L}(\theta, \sigma_\mathrm{jit.})
  = & \ \frac{1}{\sqrt{|2\pi\Sigma| |2\pi\Lambda|}}\times                 \\
    & \int \exp\left(-\frac{1}{2} \left((y-A\eta)\t \Sigma^{-1} (y-A\eta)
  +\eta\t \Lambda^{-1} \eta\right)\right) \mathrm{d}\eta,\nonumber
\end{align}
which can be integrated as
\begin{equation}
  \label{eq:lmargperf}
  \mathcal{L}(\theta, \sigma_\mathrm{jit.})
  = \sqrt{\frac{|2\pi C|}{|2\pi\Sigma| |2\pi\Lambda|}}
  \exp\left(-\frac{1}{2} \left(y\t \Sigma^{-1} y - x\t C x\right)\right),
\end{equation}
with
\begin{align}
  C^{-1} & = \Lambda^{-1} + A\t \Sigma^{-1} A,\nonumber \\
  x      & = A\t \Sigma^{-1} y.
\end{align}
This expression could be further simplified using the Woodbury Identity
to obtain a simple Gaussian distribution for $y$ \citep[e.g.,][]{Luger2017linear}
\begin{equation}
  \mathcal{L}(\theta, \sigma_\mathrm{jit.})
  = \frac{1}{\sqrt{|2\pi S|}}
  \exp\left(-\frac{1}{2} y\t S^{-1} y\right),
\end{equation}
where
\begin{equation}
  S = \Sigma + A\Lambda A\t.
\end{equation}
However, while this latter expression is more compact,
evaluating the marginal likelihood using it requires to compute the determinant of $S$
and to solve for $y\t S^{-1} y$,
with $S$ a $(n\times n)$ matrix.
Thus, the cost of a likelihood evaluation typically scales as $\O(n^3)$.
For $\Sigma$ and $\Lambda$ diagonal,
the matrix $S$ is actually banded with bandwidth $w=\max_k(n_{k+3}-n_{k-1})$.
This structure might allow to improve performances (scaling in $\O(w^2 n)$),
but since the number of measurements is usually much larger than the number of pieces ($n\gg N$),
the bandwidth $w$ is still a large fraction of $n$.
In such a case,
a more efficient method is to keep the marginal likelihood in the form of
Eq.~(\ref{eq:lmargperf}).
Indeed, for $\Sigma$ and $\Lambda$ diagonal,
$C^{-1}$ is a symmetric banded matrix with bandwidth 3,
and the cost of likelihood evaluations can be reduced to $\O(n)$.
We first define
\begin{align}
  u      & = \frac{y}{\tilde{\sigma}},\nonumber                                    \\
  G      & = \frac{1}{\tilde{\sigma}} * A = \frac{m}{\tilde{\sigma}} * B,\nonumber \\
  \gamma & = \frac{1}{\tilde{\sigma}} * \alpha = \frac{m}{\tilde{\sigma}} * \beta,
\end{align}
with $\tilde{\sigma} = \sqrt{\mathrm{diag}(\Sigma)} = \sqrt{\sigma^2 + \sigma_\mathrm{jit.}^2}$.
With these new notations, we have
\begin{align}
  C^{-1} & = \Lambda^{-1} + G\t G,\nonumber \\
  x      & =  G\t u.
\end{align}
We additionally introduce the piecewise notations
\begin{align}
  G^{(k)} & = \begin{pmatrix}
                \gamma_{n_{k-1}+1,1} & \dots  & \gamma_{n_{k-1}+1,4} \\
                \vdots               & \vdots & \vdots               \\
                \gamma_{n_k,1}       & \dots  & \gamma_{n_k,4}
              \end{pmatrix},\nonumber \\
  u^{(k)} & = \begin{pmatrix}
                u_{n_{k-1}+1} \\
                \vdots        \\
                u_{n_k}
              \end{pmatrix},
\end{align}
and for each piece we compute the $(4\times 4)$ matrix
\begin{equation}
  F^{(k)} = {G^{(k)}}\t G^{(k)}
\end{equation}
and the vector of size 4
\begin{equation}
  v^{(k)} = {G^{(k)}}\t u^{(k)}.
\end{equation}
We then obtain the lower banded representation of the matrix $G\t G$ with
\begin{equation}
  \label{eq:GtGband}
  \left(G\t G\right)_{i,i+d} = \sum_{j=\max(1,i-N+1)}^{\min(4-d, i)} F^{(i-j+1)}_{j,j+d},
\end{equation}
and the vector $x$ with
\begin{equation}
  x_i = \sum_{j=\max(1,i-N+1)}^{\min(4, i)} v^{(i-j+1)}_j,
\end{equation}
for $i \in [1, N+3]$ and $d \in [0,3]$.
Since $\Lambda$ is assumed to be diagonal, the lower banded representation of $C^{-1}$
is straightforward to compute from Eq.~(\ref{eq:GtGband}).
Finally, we compute the Cholesky decomposition $C^{-1} = LL\t$ in lower banded form
which allows to straightforwardly compute the determinant of $C$
and to solve for $x\t C x = (L^{-1} x)\t(L^{-1} x)$.
Using algorithms dedicated to banded matrices, both the Cholesky decomposition and the solving
have a computational cost scaling as $\O(n)$.

In practical computations, we assume that the priors on the B-spline parameters $\eta$
are sufficiently broad (i.e., the diagonal entries of $\Lambda$ are sufficiently large)
such that $C^{-1} \approx G\t G$.
Moreover, we additionally ignore the determinant of $\Lambda$ in Eq.~(\ref{eq:lmargperf})
since it is a constant renormalization factor which does not have any impact in our analyses.
These approximations are equivalent to assuming a uniform prior for the parameters $\eta$
with unspecified very large bounds.


\section{Internal structure model}
\label{ap:internal_structure} 

For both the sample of re-analysed Kepler planets and the sample of RV-characterised planets, we used a Bayesian inference method to characterise the internal structure of the planets. The full model is described in detail in \citep{Leleu2021} and based on \citep{Dorn17}.

We assume that the planets are spherically symmetric and consist of four fully differentiated layers (iron core, silicate mantle, water and H/He atmosphere). In terms of equations of state, we use \citep{Hakim} for the iron core, \citep{Sotin} for the silicate mantle, \citep{Haldemann2020} for the water layer and the atmosphere model of \citep{LopezFortney14}. {We further assume that the H/He atmosphere is independent of the rest of the planet and fix the temperature and pressure at the gas-water boundary. Thereby, we neglect effects of the gas layer on the solid part of the planet, such as atmospheric pressure or thermal insulation.}

In the sample of RV-characterised planets, for many planets at least some of the stellar observables are unknown. Therefore, we limit the stellar input parameters of the Bayesian model to the observables that are known for all planets from both samples: mass, radius and the effective temperature of the star. For these values, we assume an error of 5\%, since reliable error bars are not generally available. Furthermore, we assume the stars to be of Solar composition, also within error bars of 5\%. For the age of the stars, we assume it is unconstrained ($5 \pm 5$\,Gyr). The planetary input parameters of the model are the mass and radius values with the respective errors and the period with an assumed error of 0.1\%. Additionally, the model assumes that the composition of the modelled planet matches the one of the star exactly (see \citep{Thiabaud2015}). Note that the evidence for this is not quite conclusive and recently, \citep{Adibekyan2021} showed that the correlation might in fact not be a 1:1 relationship. 

We stress that the results from the Bayesian inference model depend to some extent on the chosen priors and would differ if very different priors were chosen. Again following the method detailed in \citep{Leleu2021}, we assume a log-uniform prior for the gas mass and a prior that is uniform on the simplex for the iron core, mantle and water mass fractions with respect to the solid planet. However, we assume we choose an upper limit of 50\% for the water mass fraction (see \citep{Thiabaud} and \citep{Marboeuf}). The results from our analysis are shown in Figure \ref{fig:R_Teq_gas}. 

\end{document}